\documentclass[fleqn,usenatbib]{mnras}
\usepackage[T1]{fontenc}
\usepackage{ae,aecompl}
\DeclareRobustCommand{\VAN}[3]{#2}
\let\VANthebibliography\thebibliography
\def\thebibliography{\DeclareRobustCommand{\VAN}[3]{##3}\VANthebibliography}
\usepackage[normalem]{ulem}
\usepackage{graphicx}	
\usepackage{amsmath}	
\usepackage{amssymb}	
\usepackage{mwe}
\usepackage{xspace}
\usepackage{cleveref}

\usepackage[dvipsnames]{xcolor}

\definecolor{darkgreen}{RGB}{50,150,0}

\definecolor{darkred}{RGB}{150,50,50}

\def\be{\begin{equation}}
\def\ee{\end{equation}}
\def\ba{\begin{eqnarray}}
\def\ea{\end{eqnarray}}
\def\({\left(}
\def\){\right)}

\usepackage{newtxtext,newtxmath}

\title[Non-Gaussianity of the CIB and Its Lensing]{Exploring the Non-Gaussianity of the Cosmic Infrared Background and Its Weak Gravitational Lensing}

\author[Jaemyoung Lee et al.]{
Jaemyoung (Jason) Lee,$^{1}$\thanks{E-mail: astjason@sas.upenn.edu} 
J.~Richard Bond,$^{2}$
Pavel Motloch,$^{2}$
Alexander van Engelen,$^{3}$ and 
George Stein$^{4, 5}$
\\
$^{1}$Department of Physics and Astronomy, University of Pennsylvania, 209 South 33rd St, Philadelphia, PA 19104, U.S.A.\\
$^{2}$Canadian Institute for Theoretical Astrophysics, University of Toronto,
60 St. George St., Toronto, ON M5R 2M8, Canada\\
$^{3}$School of Earth and Space Exploration, Arizona State University,
781 Terrace Mall, Tempe, AZ 85287, U.S.A.\\
$^{4}$Berkeley Center for Cosmological Physics, University of California,
341 Campbell Hall, Berkeley, CA 94720, U.S.A.\\
$^{5}$Lawrence Berkeley National Laboratory,
1 Cyclotron Rd, Berkeley, CA 94720, U.S.A.
}

\date{Accepted XXX. Received YYY; in original form ZZZ}

\pubyear{2022}

\begin{document}
\label{firstpage}
\pagerange{\pageref{firstpage}--\pageref{lastpage}}
\maketitle

\begin{abstract}

Gravitational lensing deflects the paths of photons, altering the statistics of cosmic backgrounds and distorting their information content. We take the Cosmic Infrared Background (CIB), which provides plentiful information about galaxy formation and evolution, as an example to probe the effect of lensing on non-Gaussian statistics. Using the Websky simulations, we first quantify the non-Gaussianity of the CIB, revealing additional detail on top of its well-measured power spectrum. To achieve this, we use needlet-like multipole-band-filters to calculate the variance and higher-point correlations. Using our simulations, we show the 2-point, 3-point and 4-point spectra, and compare our calculated power spectra and bispectra to \textit{Planck} values. We then lens the CIB, shell-by-shell with corresponding convergence maps, to capture the broad redshift extent of both the CIB and its lensing convergence.  The lensing of the CIB changes the 3-point and 4-point functions by a few tens of percent at large scales, unlike with the power spectrum, which changes by less than two percent. We expand our analyses to encompass the full intensity probability distribution functions (PDFs) involving all n-point correlations as a function of scale. In particular, we use the relative entropy between lensed and unlensed PDFs to create a spectrum of templates that can allow estimation of lensing. The underlying CIB model is missing the important role of star-bursting, which we test by adding a stochastic log-normal term to the intensity distributions. The novel aspects of our filtering and lensing pipeline should prove useful for any radiant background, including line intensity maps.

\end{abstract} 




\section{Introduction}
\label{sec:intro}

Gravitational lensing is the deflection of distant photons by intervening structure. Thus far, most attention of gravitational lensing has focussed
on optical galaxies \citep{galaxy_lensing_brainerd1996} and the Cosmic Microwave Background (CMB) \citep{Lewis2006}. In both cases, the changes to the distributions of the photons on the sky have been characterized, and observations of these changes have been used to extract critical science content on the evolution of
structure \citep{weak_lensing_cosmology_hoekstra2008}. However, all extragalactic sources are gravitationally lensed by intervening large scale structure between the source and us. Upcoming intensity mapping surveys are gaining in sensitivity and  extensive multi-line intensity mapping experiments are projected to probe more than 80\% of the volume of the observable universe \citep{line-intensity_kovetz2019}. Lensing of intensity mapping has been shown to be a challenge to detect with current surveys, but next-generation surveys might be able to detect lensing signals. This seems especially true when cross-correlation with other low-redshift tracers is utilized, potentially providing tighter constraints on cosmological parameters \citep{zahn2006_21cm_lensing, pourtsidou2016_21cm_lensing, foreman2018_21cm_lensing}. 

One example of intensity mapping with a long history of theory and observations is the the Cosmic Infrared Background (CIB), the emission from dust radiating down-shifted UV/optical radiation in star-forming galaxies to lower frequencies ( < 1000 GHz). Although the emission is from galaxy-scale objects, the finite resolution of the instruments used to detect it blend much of the emission in what appears as a diffuse component, albeit with non-Gaussian CIB anisotropies. It is possible that a significant CIB can arise from high redshift if there are luminous sources and dust, but the bulk of the emission is expected over the prime galaxy forming range, taken here to be from $ z < 4.2$. The CIB traces well the clustering of galaxies throughout the evolution of the late-time universe. Its statistics were first elucidated in \citet{BCH1}, were put in terms of a ``halo'' model, with a Poissonian shot noise term and a continuous clustering term in \citet{BCH2}. These works focussed on amplitudes and 2-point statistics, with a first CIB-source map shown in \citet{laquila} and maps with the full statistics using the Peak Patch algorithm done in \citet{BMcib2} and \citet{BMcib1}, and further developed in \citet{leshouches} in the 90s, and in \citet{WebskyPaper2020} in the 2010s. See also \citet{2000A&A...360....1G}. The model used for the dust emission in these early works is of the same form as that used in the current literature, in particular as utilized in \textit{Planck} analyses, but are based on mass-ordered emission that does not fully take into account the local perturbations that can cause galaxies of low mass to burst up in star formation activity compared with more massive ones, which can sometimes be more quiescent. The CIB model we adopt in this work is akin to the \textit{Planck} CIB model \citep{Planck2013-XXX}, which was also utilized in the \citet{WebskyPaper2020} Websky approach to extragalactic background mapping. Webskys use response functions of halos to luminous emissions, the CIB being one, to light-cone halos found using the peak patch algorithm and second order Lagrangian perturbation theory. The Websky \citep{PeakPatch2018, WebskyPaper2020} CIB maps are the direct descendants of the \citet{BMcib2} and \citet{leshouches} maps of the early 90s. Webskys also give fully correlated lensing maps to complement the CIB maps, and are the basic tools used in this paper. 

As with the CMB anisotropies, the CIB 2-point correlation and related power spectrum has been studied extensively, providing a window to the  modeling of galaxy clustering at a wide band of redshifts \citep{Planck2013-XXX}. Unlike the primary CMB radiation, the underlying localized nature of the CIB sources means it is intrinsically non-Gaussian so its monopole (average) emission and power spectrum, although valuable, misses the non-Gaussian clumpiness of the CIB. It is for this reason that the CIB bispectrum, the harmonic-space equivalent of the real-space 3-point correlation function, has been measured by multiple teams,  such
as with \textit{Planck} \citep{Planck2013-XXX}, the South Pole Telescope \citep[SPT;][]{crawford_SPT_bispectrum2014}, and the Atacama Cosmology Telescope \citep[ACT;][]{coulton2018non}. Also, the bispectrum  and higher order polyspectra have been modeled analytically, and it has been shown that complementary information from these measures can constrain the halo model further than the power spectrum on its own \citep{lacasa2014non, penin2014non}.

Here, we present a formalism of analysing higher order spectra of extragalactic backgrounds through (angular or multipole) band filtering, with an application to maps of the CIB. By passing a CIB map through a contiguous series of filters defined in harmonic space by
ranges of $\ell$'s characterized by a band centre $\ell_c$ and a band-width $\Delta \ell$ then calculating higher order map-statistics such as the skewness and kurtosis in each $\ell$-band, we can quantify polyspectra that encode (albeit reduced or projected) information on the full n-point spectra in a straightforward manner. In particular, we explicitly compute the (reduced) bispectra at \textit{Planck} frequencies. This is similar in spirit to quantifying non-Gaussianity with statistics like the bispectrum-related power spectrum and the skew-spectra presented in \cite{munshi2010} and \cite{munshi2013secondary}. This method of band
filtering proves to be a simple path to the quantification of the CIB non-Gaussianity. 

Our main target in this paper is to develop a method to lens any radiation background that is made up of localized sources distributed over a broad redshift range, using the lensing associated with all mass structures below the object's redshift. Using this, we can consider the impact of gravitational lensing on CIB statistics. Although analytic calculations suggest that the gravitational lensing of the CIB does not change the CIB
power spectrum substantially \citep{Schaan2018_analytic-CIB-lensing}, we show using our Websky simulations that its non-Gaussianity can be affected considerably. Prior to this work, weak lensing of the CIB has been investigated by \citet{Schaan2018_analytic-CIB-lensing}, \citet{mishra2019bias}, and \citet{feng2019_excess_nonG}. \citet{Schaan2018_analytic-CIB-lensing} adapted the quadratic estimator \citep{hu2002reconstruction} to reconstruct the lensing of the CIB, and probed how the non-Gaussian nature of the CIB as well as its broad redshift extent biases lensing reconstruction. \citet{mishra2019bias} quantified how much the lensing of foregrounds such as the CIB biases CMB lensing estimators, finding a small but potentially non-negligible effect. \citet{feng2019_excess_nonG} found by comparing cross correlations between CIB lensing reconstructions and tracers such as the CIB at several frequencies and the CMB lensing potential for simulations and \textit{Planck} data, that \textit{Planck} CIB measurements contain excess non-Gaussianity consistent with CIB lensing. In this paper, we explicitly quantify the change in the CIB 3 and 4-point functions due to lensing relative to the unlensed CIB statistics. 

Similar to the works of \citet{Schaan2018_analytic-CIB-lensing} and \citet{mishra2019bias}, our analysis pipeline can be adapted for various other non-Gaussian radiation fields associated source emissions, such as 21 cm, Lyman-$\alpha$, CO, CII, and other mm-wave intensity fields, to uncover further details of the 3-D structural evolution they trace.

An expansion to the Websky suite of extragalactic simulations\footnote{\url{https://mocks.cita.utoronto.ca}} \citep{PeakPatch2018, WebskyPaper2020} is used here to first show that the simulations capture the nearly-equilateral bispectra sufficiently well compared to the measured \textit{Planck} values. We then investigate the change in CIB non-Gaussianity due to lensing, accounting for the fact that the CIB sources at different redshifts are lensed by different lensing convergences. Our simulation results suggest that weak gravitational lensing can change the 3-point and 4-point functions of the CIB to measurable degrees.  This is potentially significant for observations, since the observed CIB necessarily includes gravitational lensing, while lensing has been ignored in theory predictions of the CIB bispectrum thus far.

The content of this paper is as follows. In \Cref{sec:sims}, we describe the simulations used in our study, as well  as the method we developed to accurately lens the CIB. In \Cref{sec:n-point}, we present our formalism for quantifying the n-point statistics. We then compare our simulation power spectra and bispectra with experimental data in \Cref{sec:exp_comparison}, and discuss the redshift contribution to the CIB statistics in \Cref{sec:redshift_n-point}. We present our results on the effect of lensing on the CIB statistics in \Cref{sec:lensing_npoint}. In \Cref{sec:Relative-Entropy}, we provide an alternative way to probe the effect of CIB lensing that includes additional non-Gaussian information using relative entropy. Next, we investigate stochastic effects on CIB statistics in \Cref{sec:stochastic}. We wrap up with a conclusion and discussion in \Cref{sec:concl}.

\section{Lensed CIB simulations}
\label{sec:sims}

Our methods for performing the lensing of simulated CIB maps differ from lensing of the CMB in several aspects. First, while the CMB comes from a single redshift, the CIB comes from a broad range of redshifts, and the structures hosting CIB galaxies at one redshift will lens galaxies at a higher redshift. Second, while the CMB is a diffuse, smooth field, the CIB consists of many individual sources, and we must take care to treat the flux density from each galaxy appropriately, particularly since we aim to quantify non-Gaussian properties of the CIB both before and after lensing is performed. 
In the first part of this section we briefly describe the Websky simulations \citep{WebskyPaper2020} our study is based on; we refer the reader to that paper for an in depth description. In the second part we describe how we calculate the lensed CIB signal.

The Websky simulations are based on the second order Lagrangian Perturbation Theory (2LPT) approach to nonlinear evolution of large-scale structure, 
and follow the mass-Peak Patch approach \citep{bond1996peak, PeakPatch2018} to find regions that
collapse into halos. First, peaks of density smoothed on a range of scales are flagged as
potential halo candidates. For each such candidate, the 2LPT displacement field
is then evaluated within the framework of the ellipsoidal collapse model to find regions
that will gravitationally collapse into haloes. Overlapping halo candidates are either
merged, or both kept with adjusted masses.

\begin{figure*}
    \includegraphics[width=0.96\textwidth]{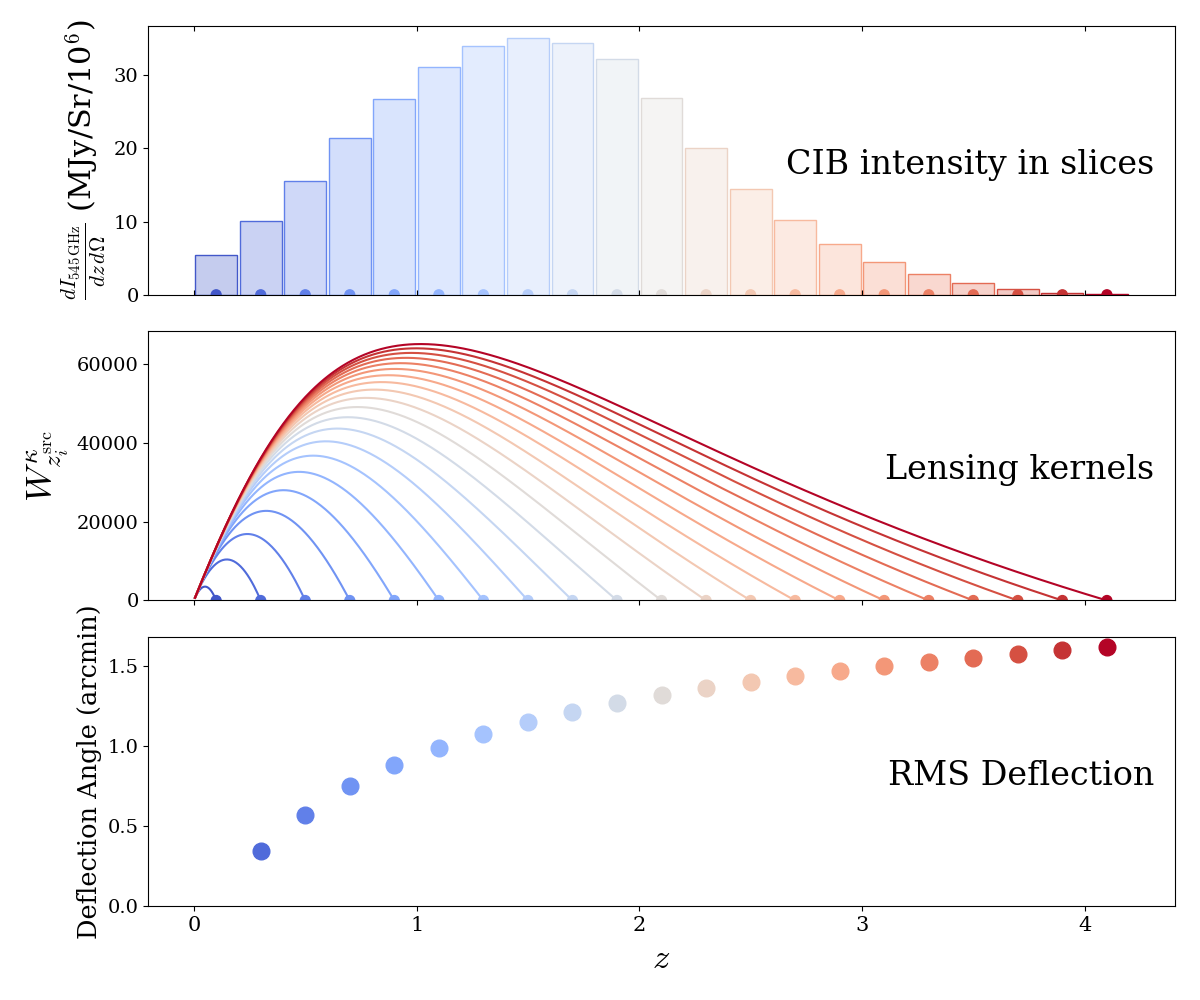}
    \caption
    {\small CIB intensity, the corresponding lensing potential kernels, as well as the RMS deflection for each redshift shell ($\Delta z = 0.2$) within our Websky CIB model. The CIB intensity increases from $z = 0$ to $z = 1.4$, peaks around $z = 1.4$ to $1.6$, then decreases until it is almost non-existent by $z = 4$. The lensing kernels, defined as $W_{z_i^{\text{src}}}^{\kappa} = \frac{3}{2} \Omega_m H_0 ^2 \frac{1+z}{H(z)}\chi(z) \big[ \frac{\chi(z_i^{\text{src}})-\chi(z)}{\chi(z_i^{\text{src}})}\big]$ with $z_i^{\text{src}}$ being the midpoint of each redshift shell in $z$-space, typically peak around half of its extent, although they display some skew towards where the CIB intensity is highest, especially as we integrate over more redshifts. The RMS deflection steeply increases at first from $0.34\arcmin$ for the first lensed shell, up to $1.7\arcmin$ for the last shell. 
    } 
    \label{Fig:slice_lensing}
\end{figure*}

\begin{figure*}
    \includegraphics[width=0.99\textwidth]{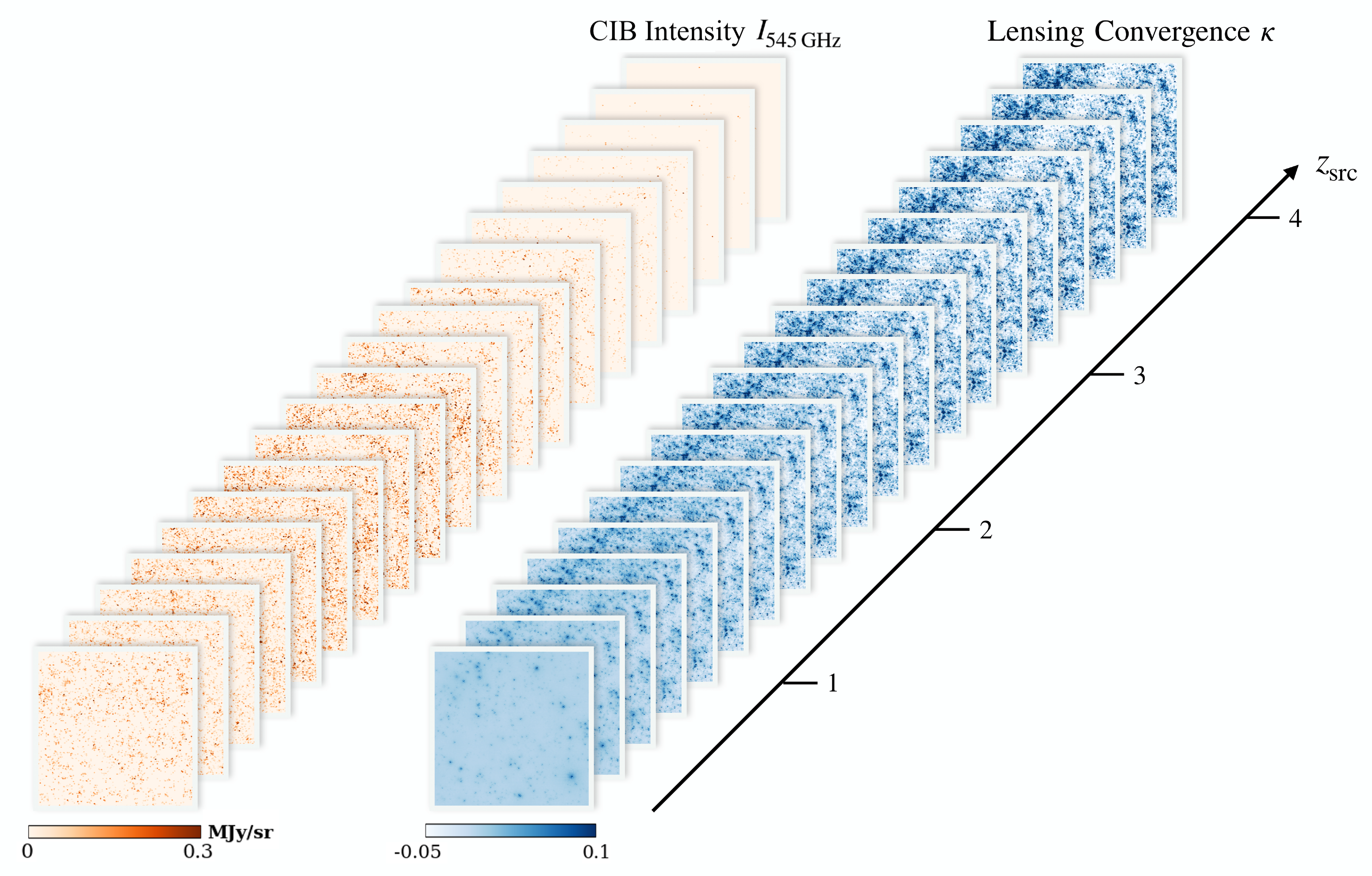}
    \caption
    {\small A patch of unlensed CIB and its corresponding lensing convergence for each redshift shell $\Delta z = 0.2$. In our simulations, each unlensed CIB shell is lensed by a convergence shell to create lensed CIB shells, which are then summed up to produce the total lensed CIB map. This method mitigates the `self-lensing' effect substantially. Note that the CIB intensity visibly thins out by $z = 3$, while the integrated lensing convergence becomes brighter at higher redshifts.}  
    \label{Fig:deck_of_cards}
\end{figure*}

Each halo is modelled as a central subhalo with a central galaxy, for massive halos, with a
number of satellite subhaloes that each hosts a single satellite galaxy; the number of subhaloes is
generated using a subhalo mass function from \citet{Jiang:2013kza}.
In the Websky simulation, there are $\sim 9 \times 10^8$ halos hosting $\sim 5 \times
10^9$ CIB galaxies. The flux density of each such galaxy is then calculated according to a CIB halo model
\citep{Shang_CIB_model/j.1365-2966.2012.20510.x}, with parameters from
\citet{viero2013hermes}. In this model, the flux density of the galaxy depends only on the mass of
its subhalo and its redshift. The key formulations of the model are given below as in \citet{WebskyPaper2020}:  

The rest-frame spectral energy distrbution (SED) of a source is given by: 
\be
\label{source_SED}
  L_{(1+z)\nu} (M, z) = L_0 \Phi(z) \Sigma(M, z) \Theta[(1+z)\nu, T_\mathrm{d} (z)]
\ee
where $\nu$ is the frequency of the observation, $M$ is the (sub)halo mass, and $z$ is the redshift. The spectral energy distribution $\Theta[\nu, T_d]$ is a greybody at low frequencies we consider for the CIB in this paper, $\Theta(\nu, z) \propto \nu^{\beta} B_{\nu} (T_\mathrm{d}(z))$ where $B_{\nu}$ is the Planck function and $\beta = 1.6$ depends on the physical nature of the dust. The effective dust temperature is given by $T_\mathrm{d} \equiv T_0 (1+z)^{\alpha}$ with $T_0 = 20.7$ and $\alpha = 0.2$. $\Phi(z) = (1+z)^{\delta_{\text{CIB}}}$ gives the redshift dependent global normalization of the $L$ - $M$ relation with $\delta_{\text{CIB}} = 2.4$ and a plateau at $z = 2$ is imposed as in \citet{viero2013hermes}. The dependence of the galaxy luminosity on halo mass is given by a log-normal function $\Sigma (M, z)$, 
\be
\label{sigma_lognormal}
   \Sigma(M, z) = \frac{M}{(2\pi \sigma_{L/M}^2)^{1/2}} \exp{[-\frac{(\log_{10}M - \log_{10}M_{\text{eff}})^2}{2\sigma_{L/M}^2}]}.
\ee
$M_{\text{eff}}$ is where the specific IR emissivity peaks, and $\sigma_{L/M}$ describes the range of halo masses that produce the luminosity. $\log(M_{\text{eff}}/M_{\odot}) = 12.3$ and $\sigma_{L/M}^2 = 0.3$ are used from model 2 of \citet{viero2013hermes}. 
$L_0$ is an overall frequency-independent normalization factor  used to scale all galaxy flux densities to match
the \textit{Planck} 545 GHz CIB power spectrum measurements \citep{Planck2013-XXX} at the angular scale
$\ell = 500$. 

This model describes a minimal parameter set that nevertheless provides a reasonable fit to the \textit{Planck} and \textit{Herschel} power spectra. One limitation is  that every galaxy at a given redshift is assumed to have the same SED, regardless of its detailed properties, including its age, merger history, and environment. In principle, it could be possible to add more parameters, but these new parameters would be largely unconstrained by current CIB data.

To calculate a map of the unlensed CIB signal, we project galaxies onto a \verb|NSIDE|
= 4096 Healpix grid and add their flux densities within each pixel. The first panel of \Cref{Fig:slice_lensing} shows the CIB intensity in each $\Delta z = 0.2$ shell for the Websky simulations. The CIB intensity peaks around $z = 1.4 \text{ to } 1.6$, which agrees with the CIB models in literature like \citet{schmidt2015_cib_z, bethermin2012_cib_model, pullen2018_CII}. There is some uncertainty in where the CIB intensity actually peaks, as some models predict the peak to be at a lower redshift \citep{schmidt2015_cib_z}, but we note that it is the star-formation rate density that we expect to peak at $2 < z < 3$ \citep{Planck2013-XXX}. We have run our analysis on a model where the CIB intensity peaks around $z = 1.1$ and found that our results are not radically altered. Although the Websky CIB maps, which go up to $z = 4.2$ as seen in \Cref{Fig:slice_lensing}, capture the essence of the CIB, we note that the CIB extends further than $z > 4.2$ in reality. 

Investigating the lensing of the CIB can be a complicated matter, especially when compared to CMB lensing. While the statistics of the unlensed CMB are very well known as it is essentially a Gaussian random field, the unlensed CIB itself is non-Gaussian and its statistics are not as well understood. Not only has the CMB power spectra been measured quite precisely \citep[most recently,][]{Planck2018_powerspectra}, the gravitational lensing effect on the CMB has been detected, and the matter between the observer and the CMB has been reconstructed to high precision \citep{planck2018_VIII}, utilizing quadratic estimators like the one given in \citet{okamoto2003_lensing_reconstruction}. As mentioned in \Cref{sec:intro}, the CIB power spectra and bispectra have been measured by a number of experiments, but the bispectrum measurements tend to have large uncertainties and its lensing reconstruction is in relatively early stages of development. \citet{Schaan2018_analytic-CIB-lensing} discuss methods to mitigate biases when reconstructing the lensing mass fluctuations from maps of the CIB. In particular, challenges arise both because of the intrinsic non-Gaussianity of the CIB itself, as well as the fact that there is redshift overlap between the lensing mass and the emissive sources. Such complications provide a motivation for our study using simulations.

Unlike the CMB, which is sourced at a narrow range of redshifts around $z \sim 1100$ at the surface of last scattering, CIB
sources are spread across a wide range of redshifts between $z = 0$ and $z = 4.5$.
Using the full 3-D information from the Websky simulation, we generate a lensing convergence map for each redshift slice of CIB galaxies.  This is a more involved treatment than that of  \citet{Schaan2018_analytic-CIB-lensing}, who took all the lensing matter to be at effectively a single redshift.
Specifically, we split the galaxies into 21 redshift
shells of width $\Delta z = 0.2$, and each galaxy in the $n$-th redshift shell is lensed by
all the matter in the first $n-1$ shells\footnote{We thus explicitly neglect lensing
effects between galaxies within the same redshift shell, which is is a reasonable approximation due to the form of the redshift-dependent lensing kernel.}. This way we correctly account
for the time evolution of the lensing effects and the fact that depending on the situation
an individual galaxy can count as source of either CIB signal or lensing. To further
simplify the calculation, when a galaxy acts as a source of CIB signal, we assume it is
located at the central redshift of its redshift shell. This allows us to pre-calculate the
lensing convergence for each redshift shell. 

For each redshift shell, the lensing calculation starts with obtaining a \verb|NSIDE| =
4096 map of the lensing convergence $\kappa$, obtained by integrating an appropriately
weighted density field along the line of sight of the central positions of the individual
map pixels.  This matches what was done for the CMB lensing field in the released version of the Websky simulation, which was similarly calculated but for a source at $z_\mathrm{src} = 1100$. Within each halo, the density field is modeled as a NFW \citep[Navarro-Frenk-White;][]{nfw1997} profile
\citep{Zhao:1995cp} and outside the haloes it is obtained from the 2LPT calculations.  
Following \citet{WebskyPaper2020}, we also include a ``field'' component, representing the lensing matter that is in halos too small to be resolved by the simulation.

For the CMB, given the lensing potential map $\phi$ and map of the unlensed CMB, the lensed map can be obtained by evaluating the CMB at the deflected positions given by $\nabla \phi$, using a pixel-based interpolation scheme. This is the approach that is codified in the commonly used and publicly available code \verb|Lenspix| \citep{LensPix2005}; it is also the approach that was taken for the lensed CMB map released by the Websky team \citep{WebskyPaper2020}, in that case using the methods in the public \texttt{pixell} library \footnote{\href{https://github.com/simonsobs/pixell}{https://github.com/simonsobs/pixell}}. These interpolation-based codes have been shown to work accurately for the lensing of a diffuse field like the primary CMB that does not have significant fluctuations on scales near the pixel size.  The primary CMB has this property thanks to its very red power spectrum, further enhanced by the washing out of structures on scales of a few arcminutes by diffusion damping at the last-scattering surface. However, since CIB galaxies are unresolved on the arcminute scales of the maps we consider, they appear as point sources the size of a pixel and   \texttt{Lenspix} cannot be accurately used for this application. The smoothing of the signal
map on small scales, which is an inherent part of \texttt{Lenspix}, leads to several artifacts such as nonphysical negative intensities we must avoid. Therefore, we developed our own lensing pipeline. 

Rather than attempt to remap the given unlensed map with the deflection vectors, we opted to use the source galaxy catalogs and create entirely new CIB maps that include the effects of lensing. Given the convergence map $\kappa_{z_i}$ for a given source redshift $z_i$, which includes all the matter in the shells between the galaxy and the observer, we determine the lensed CIB intensity and position of each individual galaxy as follows:

\begin{enumerate}
  \item We deflect the galaxy's angular position by an amount given by 
  $\vec{\alpha} = \nabla \phi_{z_i}$, with the lensing potential $\phi_{z_i}$ given in terms of $\kappa_{z_i}$ by $\kappa_{z_i} = \frac{1}{2} \nabla^2 \phi_{z_i}$. In terms of the 3D gravitational potential $\Phi(\chi, \eta)$ at radial comoving distance $\chi$ and conformal lookback time $\eta$, we have
  \be
  \phi_{z_i}(\mathbf{\hat n}) \equiv 2 \int_0^{\chi({z_i})}d\chi \frac{\chi({z_i}) - \chi}{\chi({z_i})\chi}\Phi(\chi \mathbf{\hat n}; \eta_0 - \chi).
  \ee  
  More specifically, using spherical trigonometry, we invert Equations A15 and A16 of \citet{LensPix2005}, $\cos{\theta'} = \cos{d}\cos{\theta} - \sin{d}\sin{\theta}\cos{\alpha}$ and $\sin{\Delta \phi} = \frac{\sin{\alpha}\sin{d}}{\sin{\theta'}}$, where the lensed position is $(\theta, \phi)$ and the unlensed $(\theta', \phi + \Delta \phi)$ on a sphere, with the deflection vector given as $\mathbf{d} \equiv \nabla \phi_{z_i} = d_{\theta} \mathbf{e_{\theta}} + d_{\phi} \mathbf{e_{\phi}} = d \cos{\alpha} \mathbf{e_{\theta}} + d \sin{\alpha} \mathbf{e_{\phi}}$. Inverting the equations above and reversing the deflection vector allows us to compute the lensed position of the galaxy from the unlensed position and the $\kappa$ corresponding to the unlensed pixel location of the galaxy following the Born approximation.
  \item  Since the deflection we have described will not change the apparent brightness of the galaxy, which is assumed to be a point source, we then multiply the galaxy's flux density by the magnification factor $[(1-\kappa_{z_i}^2) - \gamma_{z_i}^2]^{-1}$, where $\gamma_{z_i} \equiv \frac{1}{2}(\phi_{z_i,11}-\phi_{z_i,22})+i\phi_{z_i,12}$ \citep[e.g.,][]{ bartelmann2001weak}.  
\end{enumerate}

Once we have the deflected and magnified galaxy positions, we combine lensed flux densities of all galaxies in a redshift bin in each pixel to obtain the total lensed CIB intensity from one redshift bin. We then add contributions from all redshifts to obtain the full lensed CIB map.

To account for the finite pixel size, we smooth each such $\kappa$ map with a Gaussian
beam of $\sigma = (\sqrt{3}N_\mathrm{SIDE})^{-1}$, which roughly corresponds to the
effective radius of a pixel. The unsmoothed $\kappa$ maps contain a substantial number of pixels where $\kappa > 0.1$ (some are even larger than $0.3$), which no longer lies in the weak-lensing regime. By smoothing the $\kappa$ map, we are able to reduce areas with large $\kappa$ values considerably. Smoothing the $\kappa$ maps suppress the $\kappa$ power spectra by about 50\% at $\ell = 6000$. Below, we will address how this choice impacts our results.

Because there still are non-negligible numbers of pixels in the $\kappa$ maps where the pixel values are larger than $0.1$ even after the smoothing, we found that we must use the exact magnification factor $[(1-\kappa^2) - \gamma^2]^{-1}$, rather than the often-used approximation $(1 + 2\kappa)$. We use the transformation between $\kappa_{\ell m}$'s and $\gamma_{\ell m}$'s given in Eq. (11) of \citet{jeffrey2021dark} as well as \verb|healpy|'s \verb|alm2map_spin()| function to generate $\gamma_1$ and $\gamma_2$ maps. The shear factor that enters in the magnification is given by $\gamma^2 = \gamma_1^2 + \gamma_2^2$. 

We note that our methodology of lensing each CIB shell with its corresponding $\kappa$ shell does not account for the fact that the halos and ``field'' components used to generate the lensing convergences get increasingly lensed as we move out to farther redshifts, in an effect known as lens coupling. It has been shown that incorporating this effect, along with other post-Born effects, suppresses the squeezed bispectrum and enhances the equilateral bispectrum of $\kappa$, especially at $z > 1$, although the impact on the $\kappa$ power spectrum is minimal \citep{pratten2016impact,fabbian2019_lens_coupling}. 
Using lensing convergences not including some of these higher-order terms may impact our analysis, but we expect that not entirely capturing the non-Gaussianity of the lensing convergences does not significantly impact our results since the change in the CIB non-Gaussianity due to lensing is primarily caused by large scales of the lensing convergence maps. One potential way to incorporate some of the post-Born effects would be to lens each of the halo + field shell with its corresponding $\kappa$ before projecting it to the $\kappa$ shells, but this is beyond the scope of this work.

\begin{figure*}
    \includegraphics[width=0.96\textwidth]{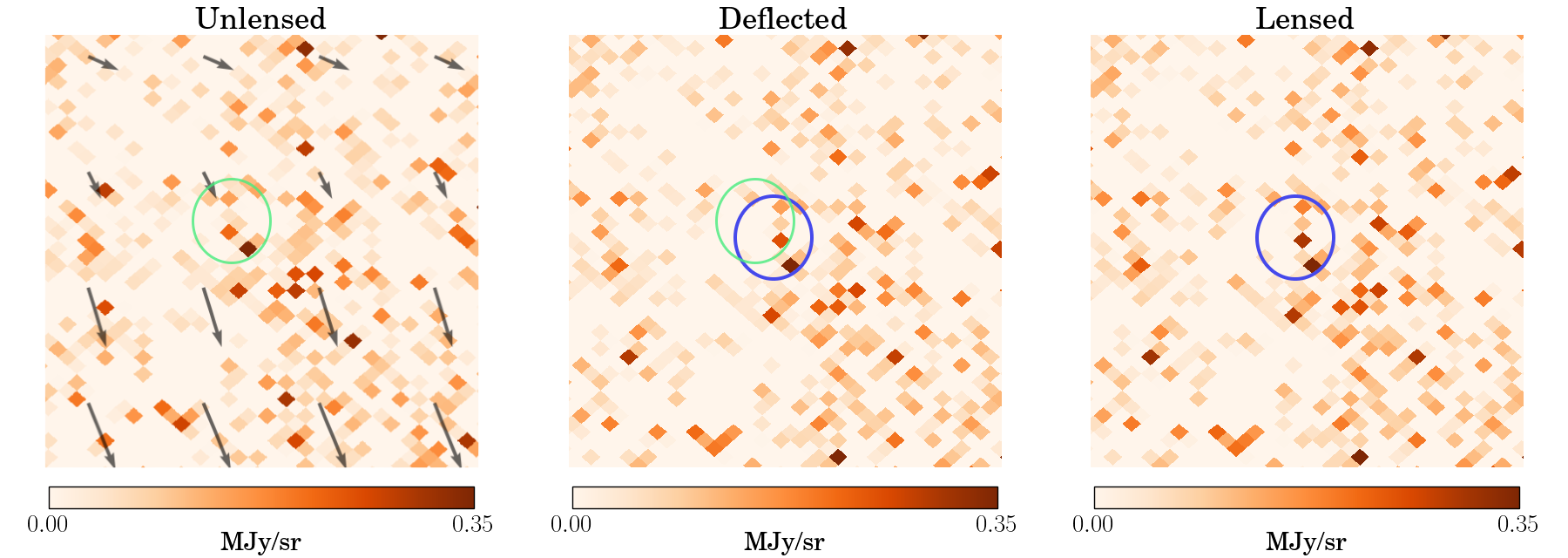}
    \caption
    {\small HealPix maps of unlensed, deflected (no magnification), and lensed CIB for a small ($0.5^\circ \times 0.5^\circ$) patch of sky
    centered on $z = 1.1$. Here, a ``lensed'' galaxy has both been deflected and has had its flux density magnified appropriately. The arrows denote the direction and magnitude of deflection. The light circled patch (Unlensed and Deflected) and the dark circled patch (Deflected and Lensed) are the same small patch of sky emphasized. One can clearly see the deflection by comparing the Unlensed and Deflected, and the magnification effect by comparing the Deflected and Lensed. 
    } 
    \label{Fig:lensing_cartview}
\end{figure*}

\begin{figure*}
    \includegraphics[width=0.48\textwidth]{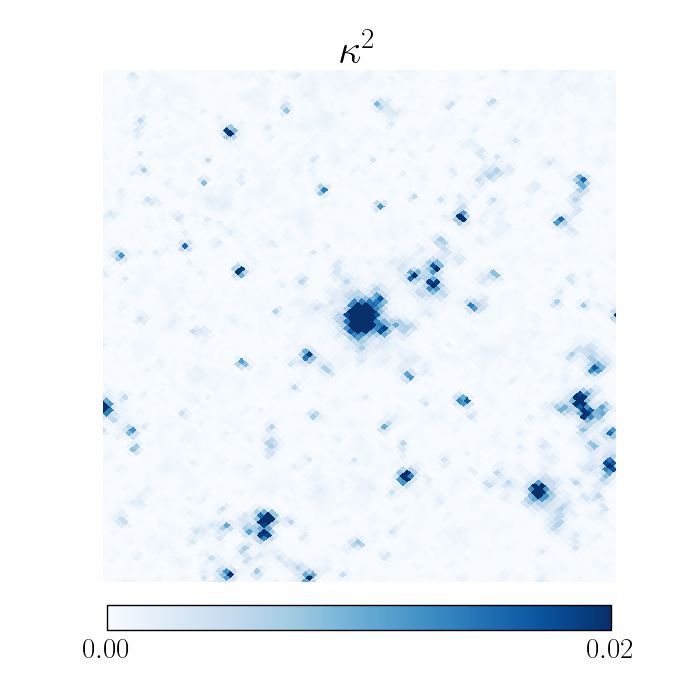}\includegraphics[width=0.48\textwidth]{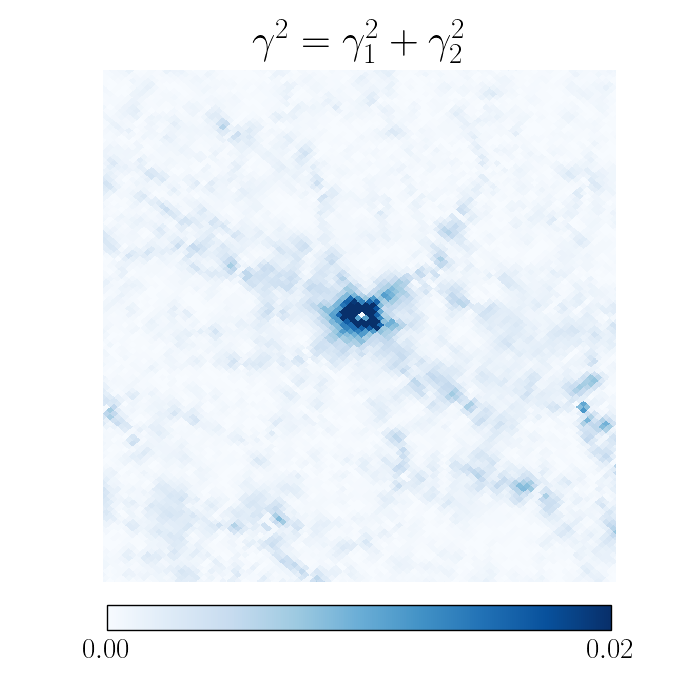}
    \caption
    {\small 
    HealPix maps of the convergence-squared ($\kappa^2$) and shear-squared ($\gamma^2 = \gamma_1^2 + \gamma_2^2$) for a ($1.5^\circ \times 1.5^\circ$) patch of sky spanning up to $z = 1.0$ (so these are from the $\kappa$ and $\gamma$ shells that lens the CIB shell shown in \Cref{Fig:lensing_cartview}). The clumps of matter are evident from the bright pixels in the $\kappa^2$ map while the $\gamma^2$ closely traces the $\kappa^2$ but the bright regions are slightly different. The cosmic web structure can be seen more clearly in the $\gamma^2$ map. 
    } 
    \label{Fig:kappa_gamma_cartview}
\end{figure*}

In \Cref{Fig:slice_lensing}, we show the total CIB intensity in each redshift shell, their corresponding lensing kernels, and the RMS deflection angle. It can be seen that the lensing kernels and CIB intensity distributions are overlapping and trace the same matter fluctuations. The RMS deflection (bottom panel of \Cref{Fig:slice_lensing}) in each redshift shell is calculated to range from 0.34 arcminutes for the very first lensed shell ($0.2 < z < 0.4$), to 1.6 arcminutes for the very last shell ($4.0 < z < 4.2$). The increase in the deflection is initally steep, but becomes more steady starting around $z = 1$, with the RMS deflection being about 1.3 arcminutes at $z = 2$. Overall, the mean RMS deflection for the CIB is around half of the amount (2.7 arcminutes RMS) that the CMB is deflected \citep{Lewis2006}. In \Cref{Fig:deck_of_cards}, we show all of the CIB shells as well as the corresponding $\kappa$ shell that ``lenses'' each CIB shell with their redshifts. It can be seen that the $\kappa$ shells are highly correlated, due to the overlapping lensing kernels, while the CIB shells are independent. \Cref{Fig:lensing_cartview} shows a $0.5^\circ \times 0.5^\circ$ patch of the CIB at 545
GHz for the redshift shell centered on $z = 1.1$. We show the unlensed, deflected (no
magnification) and fully lensed signal. One can clearly see the downwards shift due to the
deflection and the change in brightness after the magnification. In \Cref{Fig:kappa_gamma_cartview}, we display $1.5^\circ \times 1.5^\circ$ patches of the $\kappa^2$ and $\gamma^2$ maps. We note that the $\kappa^2$ shows the clumps of matter inside the halos, while $\gamma^2$ shows the regions that surround the halos to have relatively high $\gamma^2$ values. 

\section{N-point correlation functions with harmonic band filters}
\label{sec:n-point}

Our main tool to study the effect of lensing on statistics of the CIB, especially its
non-Gaussianity, will be the evaluation of variance, skewness and kurtosis of maps band
limited to a certain range of spherical harmonic coefficients $\ell$. As we show here, the
former two are directly related to the power spectrum and equilateral bispectrum of the
CIB map.

Given a CIB map $I_{\nu}$ at frequency $\nu$ and its expansion into spherical harmonic coefficients 
\be
  I_{\nu}(\hat{n}) = \sum_{\ell m} a_{\ell m} Y_{\ell m}(\hat{n}), 
\ee
we can define a band-filtered CIB map by only considering the $a_{\ell m}$'s within a
particular range of $\ell$, where $\ell_c$ is the center of an $\ell$-band,
\be
  I_{\nu}^{\ell_c}(\hat{n}) = \sum_{\ell = \ell_c - \Delta \ell/2}^{\ell_c + \Delta/2}\sum_m a_{\ell
  m} Y_{\ell m}(\hat{n}) .
\ee
We use $\Delta \ell = 640$ except for  \Cref{sec:exp_comparison}, where we compare with \textit{Planck}'s
experimental data and use their binning of $\Delta \ell = 128$.

Given filtered maps, we can calculate their variance, skewness and kurtosis as
\ba
  S^{\ell_c}_2 &=& \langle ( \Delta I_{\nu}^{\ell_c}(\hat{n}) )^2\rangle,\\
  S^{\ell_c}_3 &=& \langle ( \Delta I_{\nu}^{\ell_c}(\hat{n}) )^3\rangle\mathrm{, and}\\
  S^{\ell_c}_4 &=& \langle ( \Delta I_{\nu}^{\ell_c}(\hat{n}) )^4\rangle - 3 \(S_2^{\ell_c}\)^2 ,
\ea
where the average $\langle . \rangle$ is an average over map pixels, and $\Delta I_{\nu}^{\ell}(x)$ has the mean of the map subtracted. The subtraction in the last
line ensures that $S_4^{\ell_c}$ is zero for a Gaussian random field. Notice alternative definitions can be found in the literature \citep[e.g.,][]{Moments_Ben_David2015}.

We checked that the sharp edges of the top-hat filters do not cause significant anomalies by
running our maps through top-hat filters apodized at the edges with a sine-function as given in Eq. \eqref{Eq:apod_filter}, 
\be
  F(x) = \frac{1}{2}(1 \pm \sin{\frac{\pi(x - x_{\text{edge}})}{\Delta}}), \quad{x \in [x_{\text{edge}} - \Delta, x_{\text{edge}} + \Delta]}
  \label{Eq:apod_filter}
\ee
where the $(+)$ sign is for the left edge of the filter while the $(-)$ sign is for the right edge of the $\ell$ band, $x_{\text{edge}}$ is either edge of the top-hat filter, and $\Delta$ is the width of apodization. With bandwidth $\Delta \ell = 128$ (\textit{Planck} bins) and $\Delta = 5$ to $15$, we confirmed that the results are not qualitatively changed.


It is straightforward to show that the variance of the filtered map is related to the
map's power spectrum through
\be
\label{Cl_var}
  S_2^{\ell_c} = \sum_{\ell = \ell_c - \Delta \ell/2}^{\ell_c + \Delta \ell/2}
  \frac{2\ell+1}{4\pi}C_\ell .
\ee
This equation can be inverted to give a prescription for calculating an estimate for the power spectrum in the bands, 
\be
\label{eq:var_to_cl}
  \hat{C}_{\ell_c} \approx S_2^{\ell_c}
  \(
  \sum_{\ell = \ell_c - \Delta \ell/2}^{\ell_c + \Delta \ell/2} \frac{2\ell+1}{4\pi}
  \)^{-1}
  .
\ee

\noindent As usual, we define the angular bispectrum
\be
  B_{mm'm''}^{\ell \ell'\ell''} \equiv \langle a_{\ell m}a_{\ell' m'} a_{\ell''
  m''}\rangle ,
\ee
its angle-average as
\be
  B_{\ell \ell' \ell''} \equiv \sum_{m m' m''} 
  \begin{pmatrix} 
  \ell & \ell' & \ell''  \\
  m & m' & m''
  \end{pmatrix} 
  B_{mm'm''}^{\ell \ell'\ell''}, 
\ee

and the reduced bispectrum $b_{\ell\ell'\ell''}$ using 
\be
  B_{\ell \ell' \ell''} = \sqrt{\frac{
  (2\ell+1)
  (2\ell+1)
  (2\ell+1)
  }{4\pi}}
  \begin{pmatrix} 
  \ell & \ell' & \ell''  \\
  0 & 0 & 0
  \end{pmatrix} 
  b_{\ell \ell' \ell''} ,
\ee
where the matrices are Wigner-$3j$ symbols. With these definitions, we can relate the
skewness of the filtered map to $b$ as follows \citep{Komatsu_bisepctrum2001}:
\ba 
    S_3^{\ell_c}  &=&
    \sum_{\ell   = \ell_c - \Delta \ell/2}^{\ell_c + \Delta \ell/2}
    \sum_{\ell'  = \ell_c - \Delta \ell/2}^{\ell_c + \Delta \ell/2}
    \sum_{\ell'' = \ell_c - \Delta \ell/2}^{\ell_c + \Delta \ell/2}
    K_{\ell\ell'\ell''}
    b_{\ell \ell' \ell''},
    \label{eq:S_3_stat}
\ea
where
\ba
    K_{\ell\ell'\ell''} = \frac{(2\ell + 1)(2\ell' + 1)(2\ell'' + 1)}{16\pi^2}
    \begin{pmatrix}
    \ell & \ell' & \ell''  \\
    0 & 0 & 0
    \end{pmatrix}^2 .
\ea
We can invert Eq. \eqref{eq:S_3_stat} to evaluate an estimate of the equilateral bispectrum from the skewness of a
band-filtered map as
\ba 
    \hat b_{\ell_c,\ell_c,\ell_c} = 
    S^{\ell_c}_3  \(
    \sum_{\ell   = \ell_c - \Delta \ell/2}^{\ell_c + \Delta \ell/2}
    \sum_{\ell'  = \ell_c - \Delta \ell/2}^{\ell_c + \Delta \ell/2}
    \sum_{\ell'' = \ell_c - \Delta \ell/2}^{\ell_c + \Delta \ell/2}
    K_{\ell\ell'\ell''}
    \)^{-1}.
    \label{eq:skew_to_bl}
\ea 
This is analogous to the equilateral binned bispectrum estimator in \citet{bucher2016_binned_bispec}. 
We note that when $\ell_c \gg \Delta \ell$, Eq. \eqref{eq:skew_to_bl} reduces to: 
\be
    \hat b_{\ell_c,\ell_c,\ell_c} \approx 2\sqrt{3} \pi^3 S^{\ell_c}_3 (\Delta \ell)^{-3} \ell_c^{-1} \propto  S^{\ell_c}_3 \ell_c^{-1}
\ee     
where we used the approximation for the Wigner-$3j$ symbol  \citep{Bhattacharya2012_SZ_bispectrum} that 
\be
    \begin{pmatrix}
    \ell & \ell' & \ell''  \\
    0 & 0 & 0
    \end{pmatrix} \approx \sqrt{\frac{2}{\pi}} \frac{(-1)^{L/2}}{[(L-2\ell)(L-2\ell')(L-2\ell'')]^{1/4}}
\ee
if $L = \ell + \ell' + \ell'' $ is even and zero for odd $L$. 

The methods described above are computationally inexpensive when using full-sky maps at
\verb|NSIDE| = 4096, with each run of passing a map through a series of top-hat filters and
calculating $b_{\ell\ell\ell}$ from Eq. \eqref{eq:skew_to_bl} taking about 8 minutes on one node of the Niagara cluster \footnote{\url{https://docs.computecanada.ca/wiki/Niagara}}.

As values of the CIB trispectrum are yet to be published to our knowledge, we use the kurtosis divided by $\ell_c^{2}$ ($\ell_c^{-2}S_4^{\ell_c}$), which is proportional to the equilateral trispectrum at large $\ell_c$, as a proxy
for how strongly gravitational lensing affects CIB 4-point functions.

\section{Comparison with experimental data}
\label{sec:exp_comparison}

\begin{figure*}
    \centering
    \includegraphics[width=0.99\textwidth]{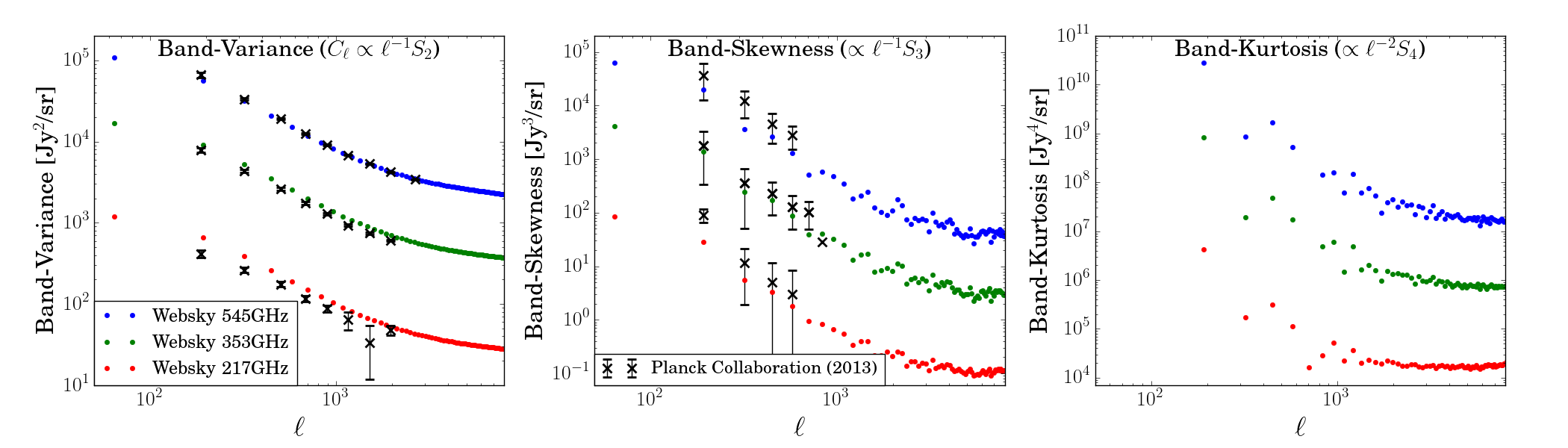}
    \caption{
    Statistics of the unlensed CIB maps from the Websky simulation at the three \textit{Planck} frequencies; top (blue) is 545 GHz, middle (green) is 353 GHz, and bottom (red) is 217 GHz. As we go to higher-order statistics, the Poisson regime becomes more evident as the spectra flatten out at $\ell > 1000$. We note that the Websky bispectra are mostly within \textit{Planck} error bars even though only the power spectra were fit to match those of \textit{Planck's}. While we do not plot the error bars for Websky values as there is only one realization, one can estimate the level of uncertainty from the scatter especially for the bispectra and kurtosis. 
    }
    \label{fig:CIB_stats_ul_Planck}
\end{figure*}

A number of surveys, including the Balloon-borne Large Aperture Submillimeter Telescope \citep[BLAST;][]{viero2009blast}, \textit{Herschel}/SPIRE \citep{amblard2011spire}, ACT \citep{dunkley2011atacama}, \textit{Planck} \citep{Planck2013-XXX} and SPT 
\citep{hall2010spt_cl, SPT2021_Reichardt, crawford_SPT_bispectrum2014}, have measured the CIB
power spectra and bispectra. In this section, we compare the statistics of unlensed CIB maps obtained from Websky with some of these experimental results.

The \textit{Planck} team measured the CIB power spectra up to $\ell = 2000$ and bispectra up to $\ell = 800$. 
We focus on the three frequencies where \textit{Planck} has both power spectrum and bispectrum measurements (217 GHz, 353 GHz and 545 GHz); 
the corresponding experimental data are shown in \Cref{fig:CIB_stats_ul_Planck}. 

Data analysis by \citet{Planck2013-XXX} included masking bright sources over a brightness
threshold (225 mJy for 217 GHz, 315 mJy for 353 GHz and 350 mJy for 545 GHz). For a fair
comparison, we thus also mask bright sources. Because the full analysis including
incomplete sky coverage is rather involved, we decided to instead replace the pixels
brighter than the corresponding experimental cutoff with the mean of the CIB map. We
believe this treatment is sufficient, due to the rareness of the very bright sources.  When we instead replace the pixels above the
experimental cutoff with double the map mean, our results are not significantly changed.

After treating the bright pixels in this manner and calculating the power spectra and
bispectra of the CIB maps according to Eqs. \eqref{eq:var_to_cl} and \eqref{eq:skew_to_bl}, we get
results shown in \Cref{fig:CIB_stats_ul_Planck}. We also show $S^{\ell_c}_4$ results for
completeness.

As in \citet{WebskyPaper2020}, we see good agreement with the \textit{Planck} CIB power spectra, though one has to keep in mind we
scaled the amplitude of the 545 GHz power spectrum to agree with \textit{Planck} at $\ell = 500$. 
On the other hand, the comparison with \textit{Planck} CIB bispectra is a nontrivial test of our model.
While the Websky bispectra are systematically below the measured values, we generally
agree within the error bars. We also see the transitions from the clustering regime to the Poisson noise dominated regime at high $\ell$ for all three polyspectra. \textit{Planck} measurements primarily capture the clustering regime, especially for the bispectrum, while the Websky simulations also yield predictions for smaller scales.

To check the high-$\ell$ limit of our bispectrum calculations, we compare with the SPT results
\citep{crawford_SPT_bispectrum2014}. In the limit of high $\ell$, the statistics are
dominated by the 1-halo contribution and the bispectrum converges to a constant (see e.g.
\Cref{fig:CIB_stats_ul_Planck}). For the SPT flux density cut (flux cut from hereon) of 22 mJy, this constant is measured to be
\citep{crawford_SPT_bispectrum2014}
\be
  b^\mathrm{SPT}_\mathrm{Poisson} [\mathrm{220\ GHz}] = \(1.84 \pm 0.26\) \times 10^{-10} (\mu \mathrm{K})^3 .
\ee
This is in good agreement with the Websky-based theoretical expectation, given by the weighted sum of
the third power of the galaxy flux densities,
\be
    b^\mathrm{th}_\mathrm{Poisson} [\mathrm{217\ GHz}] = \frac{4\pi}{N_\mathrm{pix}} \sum S_i^3
    = 1.90 \times 10^{-10}
    \,
    (\mu\mathrm{K})^3,
\ee
where $S_i$ are flux densities of the Websky galaxies dimmer than the SPT cutoff and $N_\mathrm{pix}$ is the number of pixels in a \verb|NSIDE| 4096 map ($12\times 4096^2$). However, we note that the Websky $C_{\ell}$ value at $\ell = 2940$ is about $40.7 \text{ Jy}^2/\text{sr}$ for 217 GHz while the SPT value is about $(26.7 \pm 0.7) \text{ Jy}^2/\text{sr}$ for 220 GHz. This mismatch for the $C_{\ell}$ implies that the similar $b_{\text{Poisson}}$ values between Websky and SPT do not necessarily signify that the Websky CIB model explains SPT measurements well. A possible reason  for this is briefly discussed in \citet{WebskyPaper2020}; the Websky simulations use the \textit{Planck} CIB model and the CIB contribution from halos smaller than $\sim 1.2 \times 10^{12} M_{\odot}$ is not included.

\section{Redshift origin of CIB $N$-point functions}
\label{sec:redshift_n-point}

\begin{figure*}
    \centering
    \includegraphics[width=0.98\textwidth]{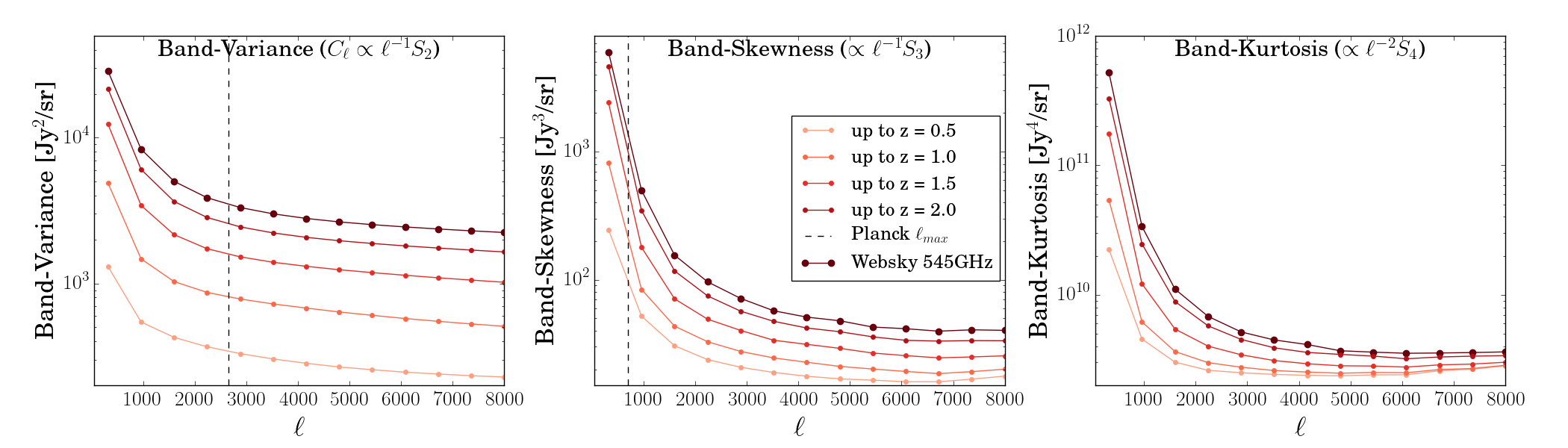}
    \caption{Contribution to the CIB statistics by redshift (cumulative) at 545 GHz. The build-up for the power spectrum and bispectrum are clearly visible. As expected, most of the CIB statistics come from $z < 2$. For the power spectrum, the contribution from each $\Delta z = 0.5$ shell remain more or less constant throughout all scales. At subsequently higher-order statistics, the contributions from $z < 0.5$ increase considerably for $\ell > 4000$.}
    \label{fig:redshift_accumulation}
\end{figure*}

\begin{figure*}
    \includegraphics[width =0.99\textwidth]{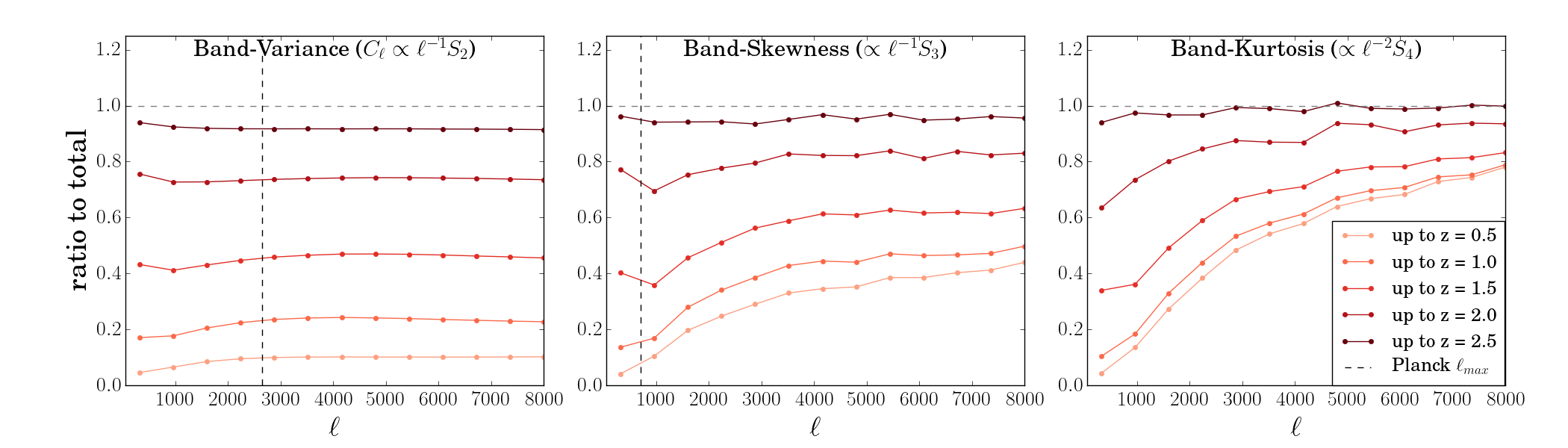}
    \caption{Cumulative-redshift ratio of the CIB statistics relative to the total CIB, showing the same information as in \Cref{fig:redshift_accumulation}. It is clear that $z < 0.5$ contributes to the CIB bispectrum and kurtosis significantly at $\ell > 4000$. However, higher redshifts are more important at low $\ell$, especially for the bispectrum in the \textit{Planck} measurement regime. Some ratios for kurtosis being larger than $1$ at high $\ell$ is the result of imposing the same (\textit{Planck}) flux cut for each cumulative shell.}
    \label{Fig:cib_stats_z_contribution_ratio}
\end{figure*}

\begin{figure}
    \includegraphics[width =8cm]{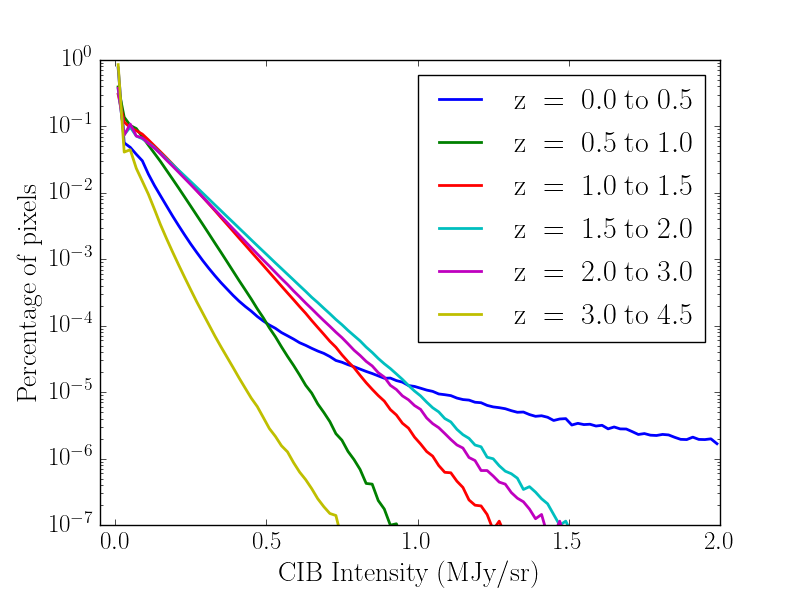}
    \caption{Breakdown of the CIB intensity histogram by redshift (no flux cut imposed). The closest redshifts contain very bright pixels from nearby galaxies, while the CIB dims significantly at $z > 3$.}
    \label{fig:cib_hist_by_z}
\end{figure}

The CIB power spectrum is dominated by the redshifts around the CIB intensity peaks and lower. For the CIB at 545 GHz, as compiled by \citet{Schaan2018_analytic-CIB-lensing} using \citet{bethermin2012_cib_model, schmidt2015_cib_z, pullen2018_CII}, the power spectrum is dominant over $0 < z < 2$ with the CIB intensity peaking around $1 < z < 2$ for various models \citep[e.g.,][]{Planck2013-XVIII, Bethermin2017, Maniyar2018, maniyar2021correlation}, whereas in the model \citet{penin2014non} adopts, the CIB power spectrum is dominated by $0 < z < 3$ with the CIB intensity peaking in the $1 < z < 3$ range. In \Cref{fig:redshift_accumulation} and \Cref{Fig:cib_stats_z_contribution_ratio}, we show the redshift accumulation of the CIB statistics as well as its ratio to the total CIB statistics. We confirm that the majority of the power spectrum comes from $0 < z < 2$ for all scales. For the power spectrum, there is only a slight scale dependence; at $\ell < 2000$, the contribution from $ z < 1.5$ decreases at lower $\ell$. For the bispectrum and kurtosis, we see a much larger contribution from $z < 0.5$, especially at $\ell > 1000$, and even moreso for the kurtosis. At $\ell > 6000$, more than 35\% of the CIB bispectrum comes from $z < 0.5$ while more than half of the kurtosis comes from $z < 0.5$ for $\ell > 4000$. While this is surprising, we attribute this to the fact that there are very bright galaxies at the closest redshifts which are not masked by the \textit{Planck} flux cut. This is evident in \Cref{fig:cib_hist_by_z}, where the \textit{Planck} flux cut of 350 mJy at 545 GHz corresponds to roughly 5.6 MJy/sr. Because the bispectrum and kurtosis are higher powers of the fluctuations by definition, they are much more sensitive to these close-by bright galaxies. We also note that \textit{Planck} has measurements with error bars up to around $\ell = 700$. \citet{penin2014non} also found that the CIB bispectra at frequencies lower than 857 GHz are dominated by low-redshift galaxies and \citet{Schaan2018_analytic-CIB-lensing} found that the CIB trispectrum is dominated by the lowest redshifts.

\section{Lensing of CIB $N$-point functions}
\label{sec:lensing_npoint}

\begin{figure*}
    \includegraphics[width =  1.03\textwidth]{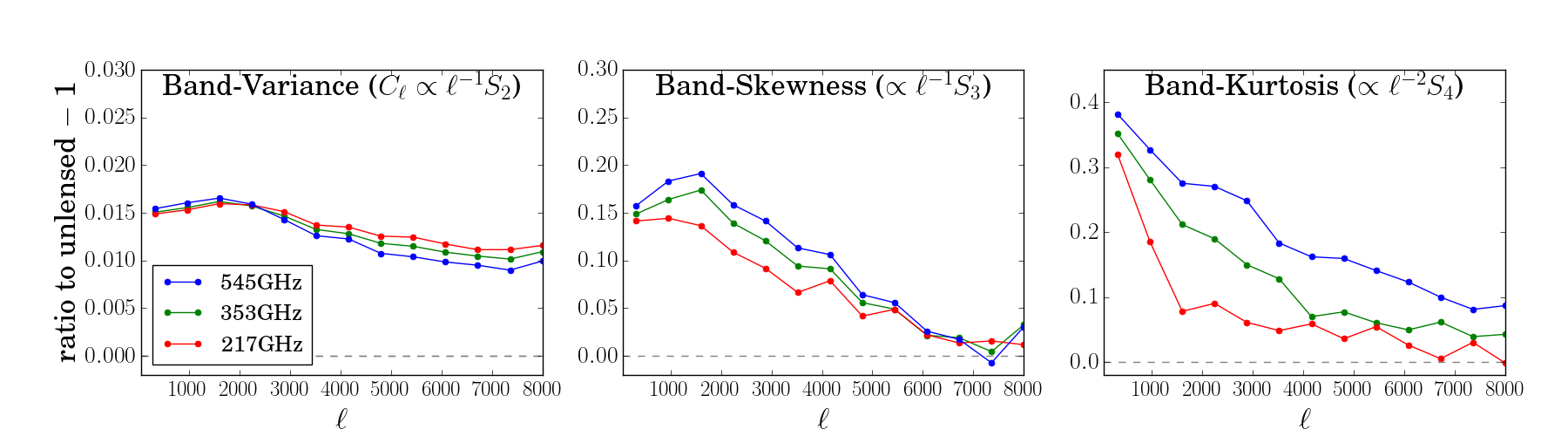}
    \caption{The effect of gravitational lensing on CIB statistics using  lensing convergence maps smoothed at the pixel level. While the power spectra are changed by less than 2\% for all three frequencies, the bispectra and kurtosis change by 10 to 40\% at low $\ell.$ The apparent discrepancy between frequencies for the bispectra and kurtosis arises due to the relatively high flux cut values for \textit{Planck} at lower frequencies.}
    \label{lensing_npoint_ratios}
\end{figure*}

In \Cref{lensing_npoint_ratios}, we show fractional changes in the CIB power spectra,
equilateral bispectra and kurtosis due to gravitational lensing.

The changes in the power spectrum are small and below 2\% at all scales, in agreement
with \citet{Schaan2018_analytic-CIB-lensing}. This is because, as visible from
\Cref{fig:CIB_stats_ul_Planck}, the CIB power spectra are relatively smooth and
featureless. Compare this with the CMB, where the significant peak structure leads to up to
$\sim$ 5\% changes in power spectra due to gravitational lensing. We do not see any
distinct difference between the three frequencies, with the peak effect around $\ell \sim
2000$ possibly related to the transition from a 2-halo to 1-halo dominated clustering
regime.

\begin{figure}
    \includegraphics[width = 8cm]{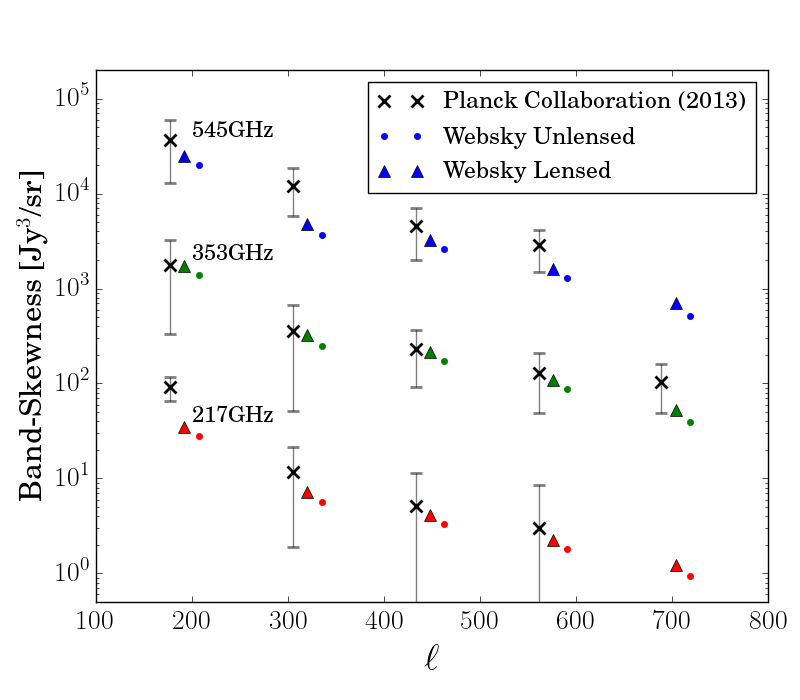}
    \caption{Comparison of unlensed and lensed Websky bispectra with \textit{Planck}
    measurements. Points are computed at the same set of central $\ell$ values but are horizontally offset for clarity. The lensed values are slightly larger than unlensed values and hence closer to \textit{Planck} values.}
    \label{fig:lensed_bl_Planck}
\end{figure}

The effects of gravitational lensing on the CIB bispectrum are more pronounced. The most
significant effect is an increase in the bispectrum by about 15\% at the largest scales, with
the lensing influence gradually dropping to small values at $\ell \sim 6000$. The effects
seem to be larger at higher frequencies. Overall, the lensing effects put our
values closer to the \textit{Planck} measurements, evident in  \Cref{fig:lensed_bl_Planck}.

The situation is similar for the 4-point function, with lensing increasing the kurtosis
of the CIB map by about 30\% at the largest scales and the lensing importance dropping as
we go to smaller scales. This time, the map frequency seems to matter much more, with the 217
GHz kurtosis showing relatively small ($< 10 \%$) lensing effects above $\ell = 1500$ and 545 GHz map
still showing $\sim$ 15\% effects at $\ell = 5000$. This however, may be attributed to the fact that the very first CIB shell spanning $0 < z < 0.2$, where there is a significant contribution for the kurtosis at high $\ell$, is not lensed in our analysis, together with the relatively high \textit{Planck} flux cuts values for lower frequencies. 

\begin{figure*}
    \includegraphics[width =0.98\textwidth]{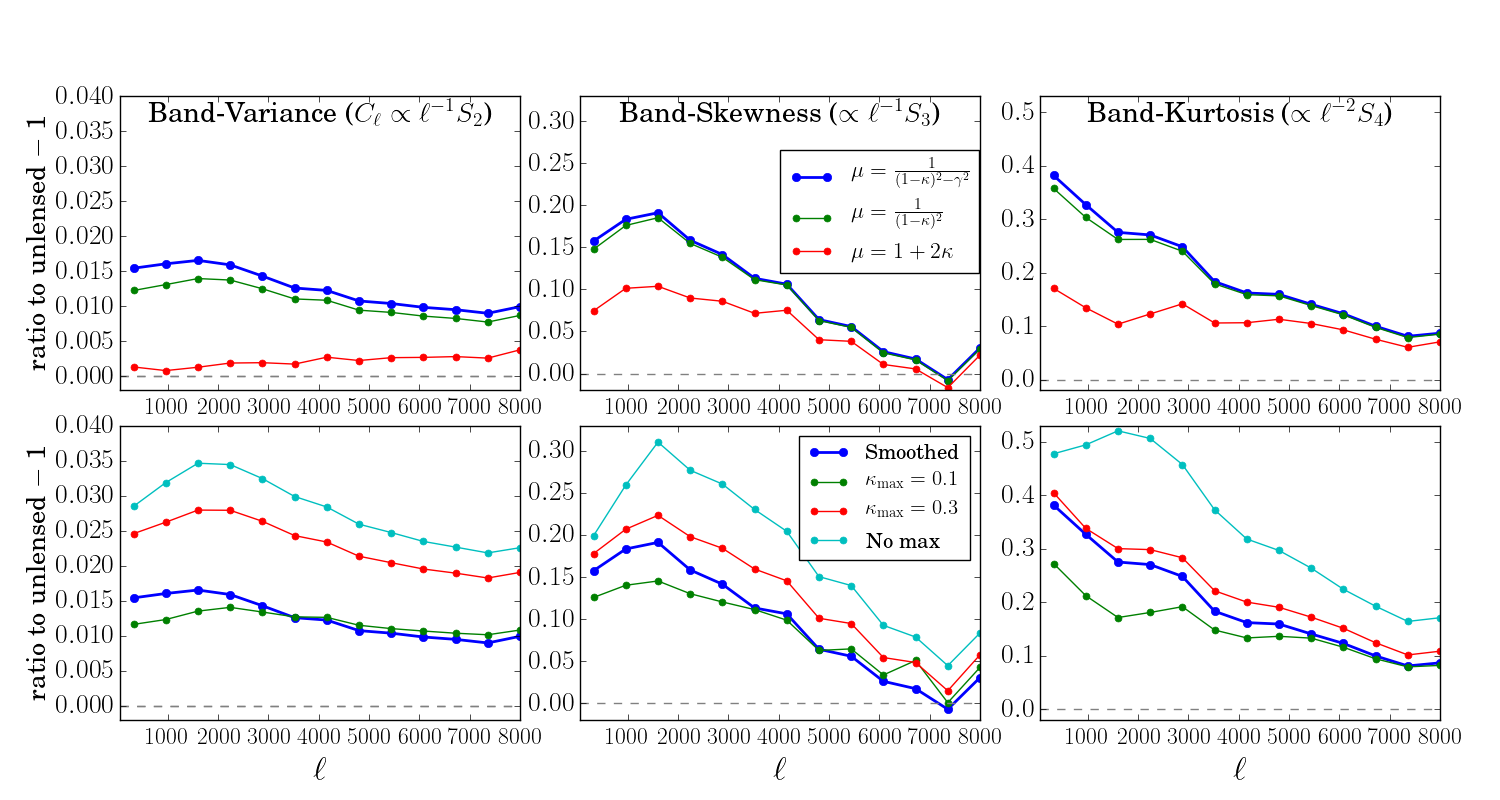}
    \caption{Effect of lensing with various methods on CIB statistics at 545 GHz, illustrating our choice of methodology. The top figures show the changes in the variance (left), skewness (middle),
    and kurtosis (right) due to lensing with pixel-level smoothed $\kappa$ maps for proper and
    approximate magnifications. The bottom figures show the changes with $\mu =
    \frac{1}{(1-\kappa)^2 - \gamma^2}$ for different treatments of the $\kappa$ maps; no treatment, setting a hard $\kappa_{\max}$, and smoothing at a pixel level. We see that the
    magnification significantly increases all 3 statistics, while using the approximation
    underestimates the non-Gaussian statistics in particular, and that setting a
    $\kappa_{\text{max}}$ or smoothing the $\kappa$ maps substantially lowers the
    statistics. Smoothing the $\kappa$ map reduces the change especially at high $\ell$. Results shown in the rest of the paper correspond to the choices in the thick blue lines.}
    \label{Fig:3x2_kappa_lensing}
\end{figure*}

In \Cref{Fig:3x2_kappa_lensing}, we investigate the sensitivity of these results to the
strongly lensed regions. In the top panel we show results when we
approximate the factor by which the galaxy flux densities are magnified as $[(1-\kappa)^{2} - \gamma^2]^{-1}
\approx 1 + 2 \kappa$ and find that regions where this approximation breaks down play crucial
roles in the large lensing effects we see above. We see that including the $\gamma^2$ term has a smaller effect, impacting mostly large scales for the bispectrum and 4-point function. In the bottom we show results with no smoothing 
of the $\kappa$ map (cyan), or an alternative treatment of the high $\kappa$ tails where
the values of $\kappa$ are capped at a certain maximum value $\kappa_\mathrm{max}$ (red and 
green). Again, we see sensitivity to how exactly the regions with high magnification are
treated. While we only show results for 545 GHz, other frequencies show qualitatively
similar behaviors.

\begin{figure*}
    \includegraphics[width =0.98\textwidth]{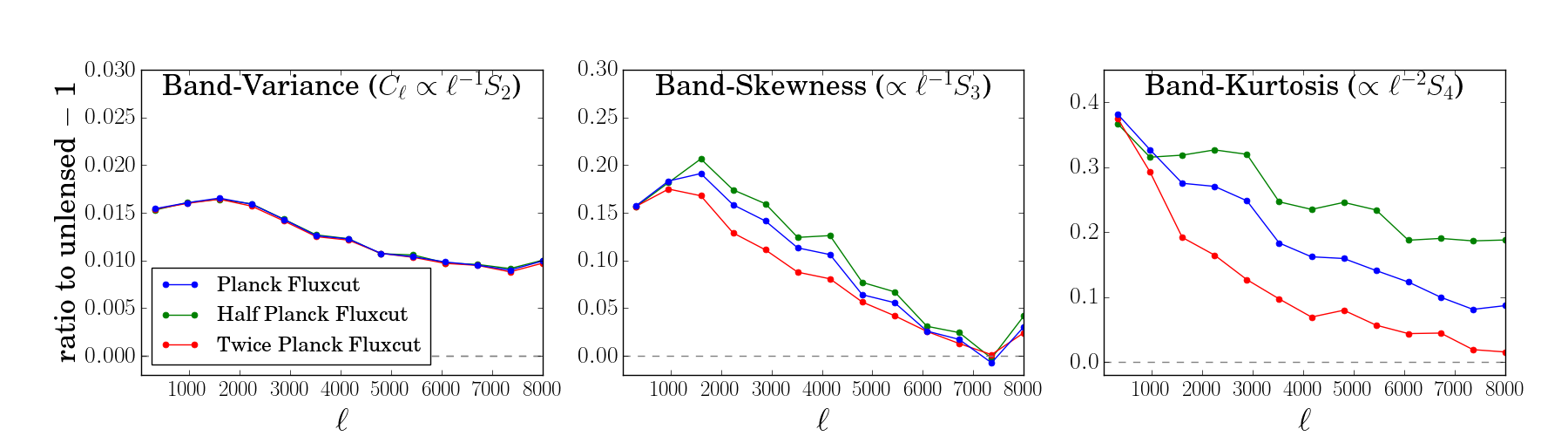}
    \caption{Effect of lensing on CIB statistics at 545 GHz with different flux cuts. Lowering the flux cut masks considerably more pixels in the very first redshift shell $z < 0.2$, which we do not lens in our analysis. This causes the  contribution to CIB statistics from $z > 0.2$ to increase, which is why the total lensing effect increases, especially for the skewness and kurtosis. Similarly, raising the flux cut decreases the lensing effect as the contribution from $z < 0.2$ becomes more important.}
    \label{Fig:lensing_fluxcut_sensitivity}
\end{figure*}

\begin{figure*}
    \includegraphics[width =0.98\textwidth]{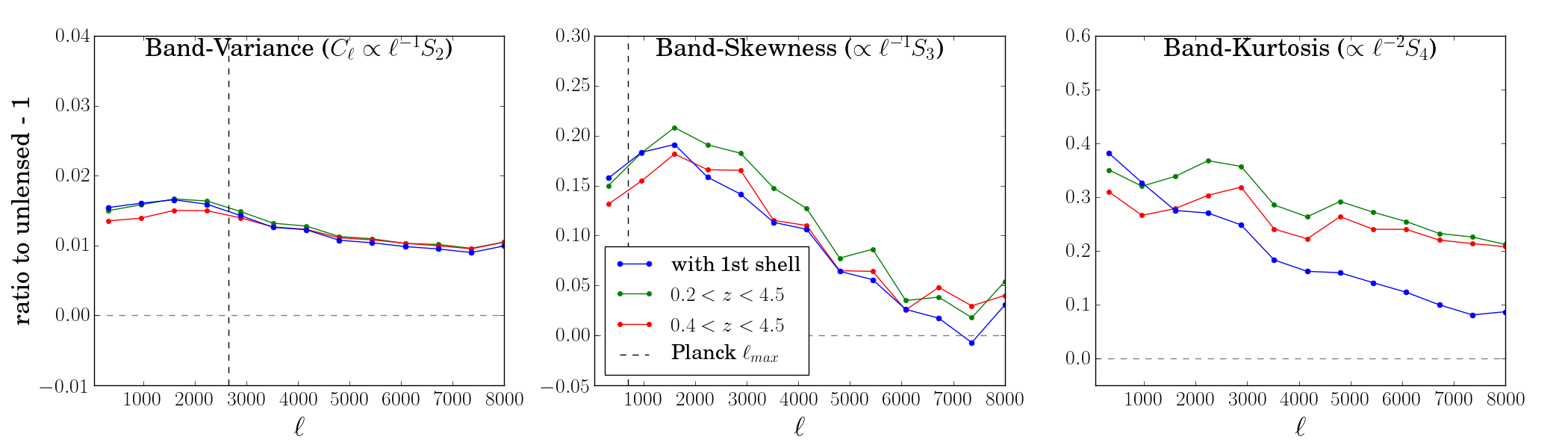}
    \caption{Effect of lensing on CIB statistics at 545 GHz with and without the very first shell $z < 0.2$, which remains unlensed. The change due to lensing is significantly more pronounced when $z < 0.2$ is excluded, especially for the kurtosis. Because the skewness and kurtosis are dominated by $z < 0.2$ starting around $\ell > 2000$, removing the first shell dramatically increases the lensing effect on those statistics.}
    \label{Fig:lensing_1st_sensitivity}
\end{figure*}

We also show how the choice of flux cuts can impact our analysis in \Cref{Fig:lensing_fluxcut_sensitivity}. For the lensed case, we impose the flux cut after the lensing, as opposed to lensing the flux cut-imposed unlensed CIB, as this is equivalent to masking bright point sources in any observations of the CIB (which are lensed). While the change in the power spectrum due to lensing is essentially the same even when the flux cut values are increased or decreased by a factor of 2, the change in the bispectrum and kurtosis are significantly altered. This is because a substantial fraction of the CIB bispectrum and kurtosis come from bright sources at $z < 0.2$ (as seen in \Cref{fig:redshift_accumulation}), and increasing or decreasing the flux cut alters the contribution of these bright sources to the higher-order statistics. Indeed, when the flux cut is lowered to half of \textit{Planck}'s, the change in the bispectrum and kurtosis increase noticeably, while the change in higher-order statistics decrease accordingly when the flux cut is increased to twice of \textit{Planck}'s. We additionally note that the effect is more pronounced for the kurtosis, a significant portion of which comes from $z < 0.2$ at $\ell > 4000$.
Because the inclusion or exclusion of the brightest galaxies in the very first shell $0 < z < 0.2$ has a major impact on the change in non-Gaussian statistics due to lensing, we further show in \Cref{Fig:lensing_1st_sensitivity} the change in CIB statistics due to lensing with and without the first shell. As expected, we find that removing the unlensed first shell altogether significantly increases the change in the bispectrum and kurtosis, especially at high $\ell$. Hence, a more accurate portrayal of the change in CIB non-Gaussianity induced by lensing
could be to break down the very first shell into finer shells (e.g. $0 < z < 0.05$ and $0.05 < z < 0.2$), and lens all but the closest galaxies, although this is beyond the scope of this paper. 

\section{Relative Entropy of $\ell$-band PDFs}\label{sec:Relative-Entropy}

\begin{figure}
    \includegraphics[width=0.47\textwidth]{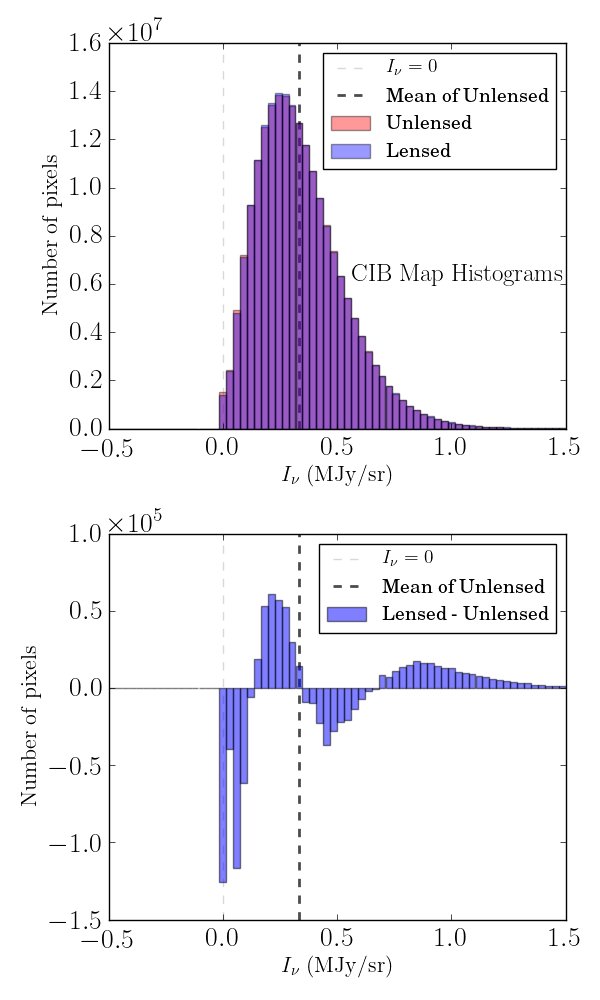}
    \caption{Top: Histograms of the unlensed and lensed Websky CIB maps at 545 GHz Bottom: The difference between the lensed and unlensed histograms, note that the number of pixels are roughly 100 times smaller for the difference histogram. While the two distributions look almost identical in the top panel, the difference of the two histograms uncover an oscillation structure in the probability distribution.}
    {\label{Fig:ul_vs_lensed_hist}}
\end{figure}

While the 3 and 4-point functions are useful descriptions of the unlensed and lensed CIB non-Gaussianity, they do not capture the full CIB non-Gaussianity. Higher order $N$-point statistics can be determined using the same method, determining the order $N$ connected components of the $\ell$-band-PDF, the probability distribution function determined in each of our $\ell$-filtered maps. Or we could work with the $\ell$-band-PDF directly, as a function of the CIB intensity. In practice, this involves constructing $\ell$-band-histograms. In \Cref{Fig:ul_vs_lensed_hist}, we show the unlensed and lensed CIB histograms at 545 GHz as well as their difference for selected bands. Although we can barely see the difference between the unlensed and lensed CIB when the two are overplotted, the difference between the lensed and unlensed clearly shows an interesting oscillation structure, caused by the deflection and hence smearing of the CIB at low-intensity pixels and magnification at high-intensity pixels. 

To quantify the effects of the lensing signature in the PDFs more explicitly, we use two relative entropy quantities, one intensive and one extensive: 
\begin{multline}
 s_{{\text{rel}}} \equiv  -\ln \frac{dN_{\text{LENSED}}}{dN_{\text{UNLENSED}}} , \\
 dS_{{\text{rel}}} \equiv  s_{{\text{rel}}}dN_{\text{LENSED}} , \quad S_{{\text{rel}}} = \int dS_{{\text{rel}}}\ .
\end{multline}
The integral $S_{{\text{rel}}}$ is the (negative) of the Kullback-Leibler ``distance'' or KL divergence between the unlensed and lensed CIB PDF, $dN_{\text{UNLENSED}}$ and $dN_{\text{LENSED}}$. 
In the following we often refer to $s_{{\text{rel}}}$ as the unweighted relative entropy; it portrays well the far tail difference of the lensed and unlensed flux distributions. The extensive $dS_{{\text{rel}}}$, which we refer to as the weighted relative entropy, damps the extreme tails by the action of the $dN_{\text{LENSED}}$ at large intensity, hence is more focussed on the mean and its vicinity, including the variance, skewness and kurtosis that we have concentrated on so far. 

With the weighted version, one has a choice of multiplying by the lensed or unlensed PDFs, destroying the anti-symmetry between lensed and unlensed. (The small difference in the two PDFs means that the asymmetry is actually also small.)  The information in the tails is visually enhanced by the logarithm, allowing for effective exploration of intensity regions far from the mean. The PDF-complexity can thus encode much more of the non-Gaussianity than the low order connected components. Further extension would consider the PDFs to vary over coarse-grained space, with PDF-PDF spatial correlation analyses. We do not develop that extension here. 

If we are trying to use relative entropy in practice, then we would be taking the observed histogram as the fully-lensed one, and would rather want to compare it with a model distribution that is lensed, but with a smaller or larger preferred amplitude than the observed PDF gives. 
We use a differential amplitude $\delta q_{\text{lens, CIB}}$ to characterize the variation of the observed from the theoretical lensed CIB: 
\begin{multline}
s_{\text{rel}} (\delta q_{\text{lens, CIB}})= - \ln dN_{{\text{obs, CIB}}}(I_{\nu },\ell_b) \nonumber \\
+\ln dN_{{\text{th, CIB}}}(\delta q_{\text{lens, CIB}}, I_{\nu} ,\ell_b)\, .
\end{multline}
Generally $\delta q_{\text{lens, CIB}}$ will have a non-zero mean and a variance about it. 
A perfect model given the data would have a mean of zero.  

In the differential between fully-lensed and lensing with a $\delta q_{\text{lens, CIB}} \ne 0$  amplitude, the observed PDF drops out. Similarly, the unlensed theoretical distribution drops out if we are interested in small $\delta q_{\text{lens, CIB}}$ variations from fully lensed. If our target is the explicit non-Gaussian nature of the PDF, as is often the case in cosmological applications, the KL distance of the lensed CIB PDF and a Gaussianized PDF with the same mean and variance are compared, and will have $ q_{\text{lens, CIB}}$ dependence in both PDFs, complicating the template when comparing with the observed, but also focussing attention on what is truly non-Gaussian. In all cases we define a differential CIB-lensing template 
\begin{multline}
\text{e}_{\text{lens, CIB}} (I_\nu , \ell_b)\equiv  [\partial s_{{\text{rel}}}/\partial q_{\text{lens, CIB}}] \ {\rm at} \ \delta q_{\text{lens, CIB}} =0, \\
{\rm hence} \ \delta s_{{\text{rel}}} \approx  \delta q_{\text{lens, CIB}} \, \text{e}_{\text{lens, CIB}}\ .
\end{multline}
Thus $\text{e}_{\text{lens, CIB}}$ is a basis vector in the differential expansion of $\delta s_{{\text{rel}}} $. 

A similar approach to lensing was used in the Arcminute Cosmology Bolometer Array Receiver (ACBAR) power spectrum analysis \citep{reichardt2009acbar}, with the template as the relative entropy of Gaussians with different power associated with different lensing amplitudes, $s_{{\text{rel}}} = \frac{1}{2} {\text{Trace}} \ln C(0)C^{-1} (\delta q_{{\text{lens, CMB}}})$, where $C(\delta q_{{\text{lens, CMB}}})$ is the CMB intensity correlation matrix at lensing amplitude $\delta q_{{\text{lens, CMB}}}$, taken relative to the fully-lensed $\delta q_{{\text{lens, CMB}}}=0$ entropy. The terminology of relative entropy was not used in those days, but it does apply to previous works such as ACBAR. As in ACBAR and \citet{calabrese2008cosmic}, and ubiquitous in all subsequent lensed CMB power spectrum work, a different multiplier was used as the lensing parameter: $A_{{\text{L,CMB}}}$, a measure of lensing strength multiplying the projected 2D gravitational potential. At the linear level the two amplitudes are proportional to each other. We could also adopt an $A_{\text{L,CIB}}$ parameterization here for the CIB, but the small differences are such that $q_{\text{lens, CIB}}$ is adequate. 

Observationally we do not know the unlensed spectrum, although much effort goes into trying to delens to isolate the unlensed. Thus the traditional expansion about unity for either $q_{\text{lens}}$ or $A_{\text{L}}$ has the classic problem of the number {\it one} in this context not being well determined since it is far from {\it zero}. Hence our emphasis here is on $\delta q_{\text{lens}}$ about {\it zero}.


The model CIB depends upon a number of parameters, which we denote by $q_{c}$, which include dust temperature, dust density, the slope of the emission, and all of their redshift dependances, and other parameters which may appear in future improved CIB models. The relative entropy can then be expanded in basis elements (linear templates) $e_c$ for each $q_c$, as well as the lensing basis element: 
\begin{multline}
\delta s_{{\text{rel}}} \approx  \delta q_{\text{lens, CIB}} \,  \text{e}_{\text{lens, CIB}}+ \sum_c \delta q_c \, e_c,  \\ 
\text{e}_{c} (I_\nu , \ell_b)\equiv  [\partial s_{{\text{rel}}}/\partial q_{c}] \ {\rm at} \ \delta q_{c} =q_{c} - q_{c,\text{current}} \, .
\end{multline}

We can then compute a \textit{spectrum} of $q_{\text{lens, CIB}}$ and $q_{c}$ in $\ell$-bands preferred by observed data. One can also combine all $\ell$-bands together since when the model is good there should be no $\ell$ dependence of the model parameters. The templates $e_c$ and $e_{\text{lens, CIB}}$ will change as the preferred parameter values change. Subsequent iteration results in best fit (final) $q_{c,\text{f}}$, and the $e_c$ dependence on $q_{c,\text{f}}$ effectively turn the iterated linear expansion into a nonlinear one, $s_{{\text{rel}}} (q_{c,f}, q_{\text{lens, CIB},f})$. 

A measure of how well the converged $s_{\text{rel}}$ does relative to the observations is the integral KL divergence, $S_{{\text{rel}}}(q_{c,f}, q_{\text{lens, CIB},f})$. If the templates are similar in shape to each other, then the associated parameters have near-degeneracies. Experimental noise can also enhance degeneracies, an issue that has been addressed in \citet{horlaville2023CII_info} for the case of $[\mathrm{C}_{\mathrm{II}}]$ line-intensity mapping. Also, with current CIB data, the high-$\ell$ regime may not be used due to the lack of experimental measurements of non-Gaussian statistics at high-$\ell$. Nonetheless relative entropy analysis of PDFs is a promising avenue. For example, it could assist future CIB lensing reconstruction studies. 

\begin{figure}
    \includegraphics[width =0.48\textwidth]{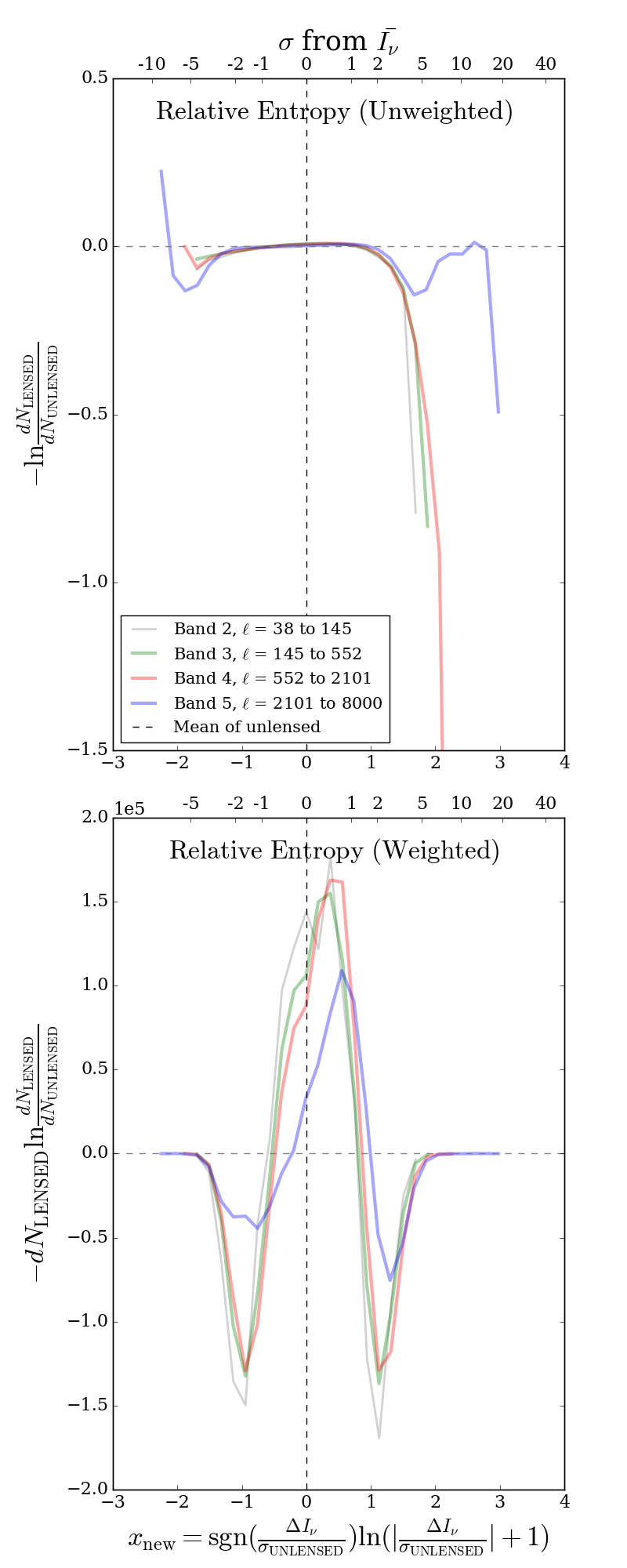}
    \caption{The full information content of CIB lensing shown using relative entropy (which are essentially log lensing templates) with smoothed $\kappa$; the top panel captures the tails while the bottom panel captures the distributions near the peaks. We show the $x_{\text{new}}$ axis, as defined in Eq. \eqref{Eq:xnew}, on the bottom and the number of standard deviations away from the mean CIB intensity on the top of each panel, and $\Delta I_{\nu} = I_{\nu} - \bar{I}_{\nu}$.}
    \label{fig:KL-entropy}
\end{figure}

In \Cref{fig:KL-entropy}, we show the unweighted and weighted relative entropies plotted against $x_{\text{new}}$, which is defined as: 
\begin{equation}\label{Eq:xnew}
    x_{\text{new}} = \text{sgn}\bigg(\frac{\Delta I_\nu}{\sigma_{\text{UNLENSED}}}\bigg) \ln\bigg(\big|\frac{\Delta I_\nu}{\sigma_{\text{UNLENSED}}}\big| + 1\bigg) .
\end{equation}
\noindent where $\Delta I_{\nu} = I_{\nu} - \bar{I}_{\nu}$ is the CIB intensity with the mean CIB intensity subtracted out and $\sigma_{\text{UNLENSED}}$ is the standard deviation of the unlensed CIB map (which is very close to that of the lensed CIB map). We use this nonlinear remapping of flux density values for the x-axis since it allows us to visualize the relative entropy more effectively at both small and large $|\Delta I_\nu|$. When $|\Delta I_\nu| \gg \sigma_{\text{UNLENSED}}$, $|x_{\text{new}}|$ approaches the logarithm of the number of standard deviations from the mean ($|x_{\text{new}}|\rightarrow |\ln{\big|\frac{\Delta I_\nu}{\sigma_{\text{UNLENSED}}}\big|}|$ as $|\Delta I_\nu| \rightarrow \infty$), whereas when $|\Delta I_\nu|$ goes to zero, the scale is linear in intensity fluctuation. We also show on top of the figures the number of standard deviations away from the map mean $\bar{I}_{\nu}$. We overplot 4 of 5 logarithmic $\ell$-bands from $\ell = 1$ to $\ell = 8000$, leaving out the first band as it spans only a few $\ell$'s. These curves can serve as the aforementioned templates for the corresponding $\ell$-bands. 


In both the top and bottom panels, we see that the curves for bands 2 ($\ell = 38 \text{ to } 145$), 3 ($\ell = 145 \text{ to } 552$), and 4 ($\ell = 552 \text{ to } 2101$) are quite similar, but not so much for band 5 ($\ell = 2101 \text{ to } 8000$). We can interpret this to mean that the lensing amplitude is consistent throughout bands 2, 3, and 4, and that the CIB intrinsic parameters assumed for the Websky simulations are plausible. The discrepancy between band 5 and the other three bands show that the smearing and magnification effects are more apparent at the smallest scales, as we expect.

The application of relative entropy of PDFs to the quantification of the amount of lensing present is done here with templates characterized by band-amplitudes. 
As elucidated earlier, a general and ambitious use of relative entropy of PDFs is to expand it in a set of templates, each with an amplitude allowing for deviation from the ``standard'' dust emission model parameters. In the CMB parlance, these could be called nuisance parameters that we wish to marginalize over to obtain the lensing amplitudes. However, these parameters are of great physical interest, and one could determine best parameters using the templates, correct the templates according to the new amplitudes obtained, iterating until the relative entropy approaches zero. Thus, not only would one show the CIB is lensed, which of course it is, but with a self-consistent spectrum one would also have a refined CIB emission model determined by the data. Though we have shown a plausible path to a full PDF analysis as a function of $\ell_b$ scale, the task on real data will be daunting. 
 
\section{Stochastic Effects}
\label{sec:stochastic}

\begin{figure*}
    \includegraphics[width = 0.98\textwidth]{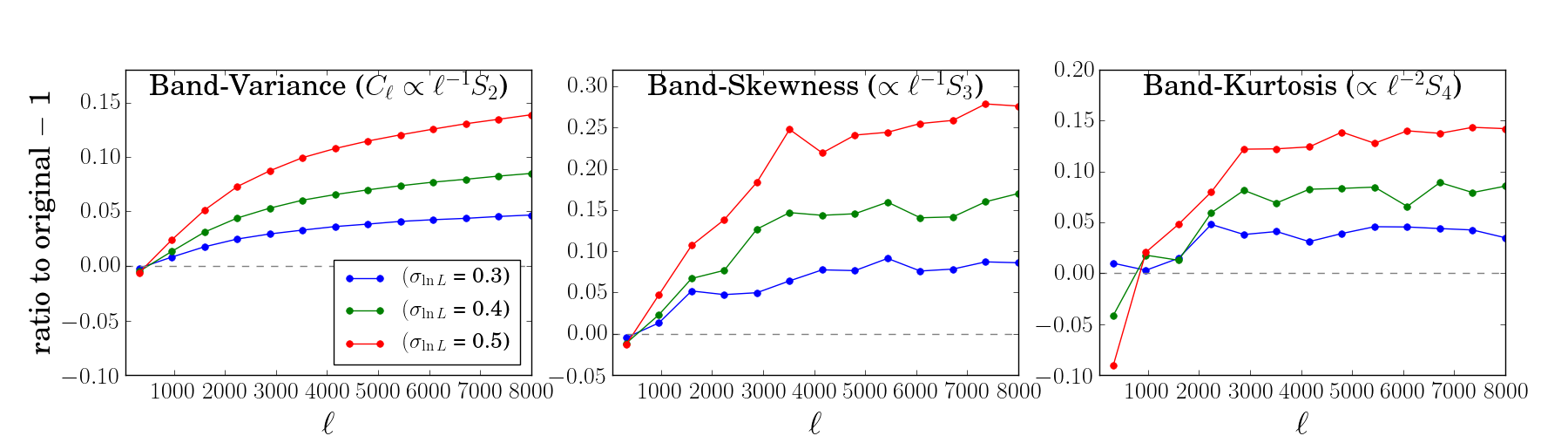}
    \caption{Effect of stochasticity on CIB statistics at 545 GHz.
     The maps were renormalized by multiplying $Ce^{-\sigma_{\ln L}^2 /2}$ with $C = [0.997, 0.994, 0.991]$ for $\sigma_{\ln L} = 0.3, 0.4, 0.5$ respectively such
    that the power spectra at $\ell = 500$ match the power spectrum of the unlensed Websky
    CIB map at 545 GHz. Stochasticity increases n-point statistics by various amounts: the increase becomes
    larger at high $\ell$ for all three statistics, but most dramatically for the
    skewness.}
    \label{fig:stochastic_n_point_ratios}
\end{figure*}

In section~\Cref{sec:lensing_npoint}, we saw that while the bispectra derived from the
Websky simulation are mostly within the \textit{Planck} experimental error bars, they are
consistently somewhat below the measurements. In this section, we test whether 
the situation can be alleviated by adding stochastic changes to galaxy flux densities. 

As mentioned in section~\ref{sec:sims}, in the CIB halo model used by Websky,  the galaxy luminosity depends only on the mass of the corresponding subhalo. In reality,
galaxy formation is affected by various environmental effects
\citep[e.g.,][]{decoratedHODs_2016/mnras/stw840} and galaxy luminosities will be consequently
altered. 

As a simple model of these stochastic effects, we multiply the
flux density of each galaxy by $\exp\left[\mathcal{N}\(0,\sigma_{\ln L}\)\right]$, where $\mathcal{N}(\mu,
\sigma)$ is a random number drawn from a Gaussian distribution with mean $\mu$ and
variance $\sigma^2$. The parameter $\sigma_{\ln L}$ allows us to control the importance of
these effects, and we consider three values: $\sigma_{\ln L} = 0.3, \text{ } 0.4, \text{and } 0.5$. More involved
models are possible, $\sigma_{\ln L}$ could be different for central and
satellite galaxies for example, but we do not consider such generalizations here.

We show changes relative to the case without the stochastic effects in
\Cref{fig:stochastic_n_point_ratios} for data at 545 GHz.
In the left panel, we see that adding these stochastic effects increases the small scale
power spectrum of the CIB at fixed large scale power. At $\ell = 5000$, we see about
3\%/7\%/12\% increase with $\sigma_{\ln L}$ of 0.3/0.4/0.5. For the bispectrum we see similar patterns, with larger relative changes as we go to
smaller and smaller scales. Effects are a bit larger than for the power spectrum, with
about 7\%/15\%/25\% increase at $\ell = 5000$ with $\sigma_{\ln L}$ of 0.3/0.4/0.5. The kurtosis shows the most interesting pattern, with a decrease at large scales and an increase at
 small scales, with the transition happening around $\ell = 1500$. The amplitude of the
change seems to increase with $\sigma_{\ln L}$ and is generally limited to below 15\%.

While inducing stochastic effects on the flux densities of CIB galaxies does not increase the bispectra at low enough $\ell$ to compensate for the difference between Websky and \textit{Planck} bispectra values (which was measured up to $\ell \approx 800$), our analysis shows that the power spectrum and bispectrum increase substantially due to stochastic effects at high $\ell$. Thus we expect that future experiments might be able to distinguish between different models of stochastic effects as they can measure the CIB power spectra and bispectra down to small scales at high precision.

\section{Conclusion and Discussion}
\label{sec:concl}

In this paper, we have provided a simple and convenient formalism for estimating the non-Gaussianity of sky maps such as the Cosmic Infrared Background.  Our method involves filtering the maps to isolate particular ranges of angular scales, and then computing the moments of the pixel distribution of the resulting map. Using this formalism, we first showed that the Websky simulations capture the equilateral bispectra of the CIB well, mostly within \textit{Planck} error bars (although slightly lower). This is remarkable because no higher-order moments, such as bispectra, were considered when constructing the Websky CIB model. We also computed the kurtosis spectrum. All three types of  spectra show clustering signals on large scales and a Poisson, or shot-noise, form on small scales. 

We then lensed the CIB using a deflection-then-magnification method, by splitting the CIB into redshift shells and lensing each shell with its corresponding lensing convergence. This method accounts for the fact that the CIB is broadly distributed over a range of redshifts. We found that gravitational lensing causes the CIB power spectra to increase by less than 2\% throughout all scales, the bispectra by 10 to 20\% at large scales, and the kurtosis by 25 to 40\% at large scales. 

We found that the change in non-Gaussian statistics are quite sensitive to several factors: \begin{itemize}
  \item the treatment of the convergence maps: setting a maximum value or smoothing; 
  \item the chosen magnification factor: using the commonly-employed weak lensing approximation $\mu \approx (1 + 2\kappa)$, rather than the more accurate $\mu \approx \frac{1}{(1 - \kappa)^2}$ or $\mu = \frac{1}{(1 - \kappa)^2 - \gamma^2}$; and 
  \item the treatment of bright, nearby sources: the chosen flux cut threshold and whether the closest redshift shell is included. 
\end{itemize}

The first two points are relevant to the fact that our calculations rely on the weak-lensing approximation, 
in which the lensing deflections are
assumed to be small. As our results show, regions where this approximation is not completely valid play
a non-negligible role in generating CIB non-Gaussianities. However, a full ray-tracing study
is beyond the scope of this initial investigation.

On top of the effect of lensing on higher-order statistics of the CIB, we also laid out a procedure for examining the full non-Gaussian information of CIB lensing using relative entropy. We provided an example of how relative entropy and its differential can be expanded using the derivatives of the differential relative entropy with respect to a parameter as basis templates. Iterating over different values of the parameter allows us to constrain the amplitude of lensing as well as the intrinsic CIB parameters. Furthermore, we explored how stochastic effects not included in the halo model could change CIB statistics, and found that the statistics increase at small scales. 

There are several reasons why this study is important. First, our lensing pipeline can be adapted for any 3-D intensity fields, such as 21cm, Lyman-alpha, as well as other mm-wave intensity fields like CO. 
In addition, CIB non-Gaussianity complicates the detection of primordial non-Gaussianity \citep{hill2018foreground,coulton2022biases}, which would be a very strong sign for inflation, and measurements of CMB lensing \citep[e.g.,][]{vanEngelen2014, osborne2014extragalactic}. To isolate these effects from intrinsic CIB non-Gaussianity, the fact that CIB lensing changes the CIB non-Gaussianity should be considered \citep[e.g.,][]{mishra2019bias}. 
Lastly, CIB non-Gaussianity provides additional information on top of the power spectrum in probing galaxy formation and clustering. In order to understand the underlying physics of galaxy formation and clustering, one needs to consider the fact that any observed CIB non-Gaussianity is lensed. We believe CIB lensing should be considered for upcoming high-precision surveys like the Simons Observatory \citep{ade2019simons} and Cerro Chajnantor Atacama Telescope-prime \citep[CCAT-prime;][]{ccat2018science, ccat2022goals}. SO is expected to observe CIB galaxies up to $z \sim 4$ at high resolution in the 280 GHz frequency band, and potentially even observe what are now galaxy clusters at earlier stages of its evolution. CCAT-prime will be able to probe higher frequencies than SO and resolve up to 40\% of the CIB at 850 GHz and detect a fraction of galaxies ($< 0.5\%$) at redshifts up to 6, providing more insight into star formation rates using individually-detected galaxies at mid-to-high redshifts than ever before. While \textit{Planck} would not have been able to distinguish the lensing from the bispectra of the CIB itself given its large error bars, not accounting for CIB lensing in such upcoming surveys could potentially cause a bias. Additionally, CIB lensing could play a role in constraining intrinsic CIB parameters through the use of lensing templates. 


\section*{Acknowledgements}

We thank Louis Pham, R\'emi Mourgues, and Joel Meyers for their contributions when this work was in its early stages. 
We thank Marcelo Alvarez for his fruitful advice regarding the Peak Patch Simulations. We also thank Fabien Lacasa for the \textit{Planck} 2013 bispectrum error bars as well as Bhuvnesh Jain and Mike Jarvis for their helpful comments on calculating the lensing effect. Additionally, we thank Giulio Fabbian for insightful comments on the higher-order statistics of the lensing convergence. We lastly thank Zack Li for the final version of the CIB catalog. 
The sky simulations used in this paper were developed by the WebSky Extragalactic CMB Mocks team, with the continuous support of the Canadian Institute for Theoretical Astrophysics (CITA), the Canadian Institute for Advanced Research (CIFAR), and the Natural Sciences and Engineering Council of Canada (NSERC), and were generated on the Niagara supercomputer at the SciNet HPC Consortium \citep{ponce2019_niagara}. SciNet is funded by: the Canada Foundation for Innovation under the auspices of Compute Canada; the Government of Ontario; Ontario Research Fund - Research Excellence; and the University of Toronto. JL acknowledges support from DOE grant DE-FOA-0002424 and NSF grant AST-2108094. AvE acknowledges support from NASA grants 80NSSC23K0747 and 80NSSC23K0464.

\section*{Data Availability}

The unlensed and lensed Websky CIB maps will be made available at \url{mocks.cita.utoronto.ca}. The lensing code is available at \url{https://github.com/motloch/cib_lensing/}.


\bibliographystyle{mnras}
\bibliography{refs}







\bsp	
\label{lastpage}
\end{document}